\documentclass[review]{elsarticle}

\usepackage{lineno,hyperref}
\modulolinenumbers[1]
\usepackage{epsfig}
\usepackage{amsmath}
\usepackage{booktabs}
\usepackage{hyperref}
\UseRawInputEncoding
\usepackage{gensymb}

\usepackage{subfigure}
\usepackage{geometry}
\usepackage{graphicx}
\usepackage{caption}

\usepackage{textcomp}
\usepackage{amssymb}
\usepackage{xcolor}
\usepackage{setspace}

\begin{document}

\begin{frontmatter}

\title{Optimization and stability analysis of the cascaded EEHG-HGHG free-electron laser}

\author[1,2]{Hanxiang Yang}
\author[3]{Jiawei Yan}
\author[1,2]{Zihan Zhu}
\author[4]{Haixiao Deng\corref{cor1}}%
\ead{denghaixiao@zjlab.org.cn}
\cortext[cor1]{Corresponding author}

\address[1]{Shanghai Institute of Applied Physics, Chinese Academy of Sciences, Shanghai 201800, China}

\address[2]{University of Chinese Academy of Sciences, Beijing 100049, China}

\address[3]{European XFEL GmbH, Schenefeld 22869, Germany}

\address[4]{Shanghai Advanced Research Institute, Chinese Academy of Sciences, Shanghai 201210 , China}

\begin{abstract}
X-ray free-electron lasers (XFELs) are powerful tools to explore and study nature for achieving remarkable advances. Generally, seeded FELs are ideal sources for supplying full coherent soft x-ray pulses. Benefiting from the high-frequency up-conversion efficiency, the cascading configuration with echo-enabled harmonic generation (EEHG) and high-gain harmonic generation (HGHG) holds promising prospects for generating full coherent radiation at 1 nm wavelength. In this paper, we design and optimize EEHG-HGHG configuration using parameters of Shanghai High-Repetition-Rate XFEL and Extreme Light Facility. In addition, we systematically analyze the effect of relative timing jitter on the output FEL performance based on various start-to-end electron beams. The intensive numerical simulations show that the cascaded EEHG-HGHG scheme can achieve 1 nm FEL pulses with peak power up to 15 GW. Further sensitivity analysis indicates that the relative timing jitter between the electron beam and seed laser has a significant impact on the FEL performance. The RMS timing jitter of 3 fs can lead to the final output pulse energy fluctuations of 29.16\%.
\end{abstract}


\end{frontmatter}

\renewcommand{\baselinestretch}{2.0}
\section{Introduction}
X-ray free-electron lasers (XFELs) are used in a wide range of frontier research in various scientific fields such as biology systems, material science and chemistry due to their ultra-fast, high-intense, and coherence {\cite{Bostedt2016,Huang2021}}. In recent years, several x-ray FEL facilities have been developed successfully or are under construction worldwide {\cite{Emma2010,Ishikawa2012,Kang2017,Decking2020,Prat2020}}. Most of these facilities utilize the self-amplified spontaneous radiation (SASE) scheme, whose radiation is initial from the shot noise of the electron beam  {\cite{Kondratenko1980,Bonifacio1994,SALDIN20021}}. The lack of temporal coherence and output intensity fluctuations limit its application in specific x-ray spectroscopy experiments. The self-seeding scheme has been proposed to improve the longitudinal coherence, which employs a monochromator to filter the SASE spectrum but cannot avoid the output fluctuations {\cite{FELDHAUS1997341}}. Seeded FELs based on electron beam manipulation utilizing external coherent seed lasers {\cite{Lewenstein1994,Yu1991,Stupakov2009,Xiang2009,Deng2013,Feng_2014}} are favorable candidates for generating full coherent soft x-ray FEL pulses accessing advanced applications such as time-resolved pump-probe absorption spectroscopy and x-ray resonant scattering experiments, etc {\cite{Huang2021}}.

As a typical seeded FEL scheme, the high-gain harmonic generation (HGHG) has been demonstrated in various FEL facilities to generate stable and coherent radiation {\cite{Yu1991,Cutic2011,Allaria2012}}. However, due to the increased energy spread in the modulation and the exponential gain requirement, the harmonic up-conversion of the single-stage HGHG is limited to 15, covering the EUV range from the external UV laser {\cite{Allaria2012}}. Another recently demonstrated seeding scheme is the echo-enabled harmonic generation (EEHG) which can obtain ultra-high harmonic components {\cite{Stupakov2009,Xiang2009}}. Owing to the nonlinear manipulation of the longitudinal phase space, the EEHG is insensitive to the electron beam imperfections {\cite{Xiang2010,Zhao2012}}. The proof-of-principle experiments have demonstrated the EEHG scheme can obtain the 75th bunching with the far-infrared laser {\cite{Hemsing2016}}. Meanwhile, the coherent FEL radiation with the wavelength of the EUV and soft x-ray has been realized using the UV seed laser {\cite{RebernikRibic2019}}. The EEHG scheme experimentally has a more stable output FEL performance than the cascaded HGHG {\cite{RebernikRibic2019}}. However, due to various three-dimensional effects, the increase of harmonic up-conversion in the EEHG scheme is challenging in practical conditions. Cascaded multi-stage HGHG has been proposed and demonstrated with the “fresh bunch” technique to increase the harmonic up-conversion number {\cite{Yu1997,Yu2003}}. To further extend to the soft x-ray region, the cascading of at least two stages of HGHG are required, which are more sensitive to energy beam imperfections, such as energy chirp, resulting in the degradation of the FEL properties {\cite{Allaria2013,Liu2013,Penn2014,Feng2019}}. Therefore, the combination of the EEHG and HGHG is the most promising configuration to generate full coherent sub-nanometer wavelength FEL pulses, where the EEHG is used as the first stage to seed the second stage HGHG with the fresh bunch technique.

During the design and optimization of the cascaded EEHG-HGHG configuration, several crucial issues should be considered: first, large slice energy spread is the most basic issue for seeding. With the rise of energy spread, a higher peak power seed laser is essential, with more stringent requirements for the state-of-art laser system {\cite{Yu2000,Penco2020}}; second, synchronization issues are critical for seeded FELs, such as the temporal and spatial overlap of the ultrashort electron beam and external seed lasers, which can directly degrade FEL performance {\cite{Danailov:14,Wang2016,yang:ipac}}; third, current spikes should be eliminated as much as possible. Because of the tens fs-scale electron beam and seed lasers, the radiation produced by the interaction of the laser and the beam cannot suppress SASE produced by the unseeded part, resulting in its early saturation and lower pulse energy, which can destroy the longitudinal coherence {\cite{Hemsing2019}}; finally, evaluation of shot noise effects from electron beam and imperfections of seed lasers is necessary {\cite{Feng2013}}. These shot noises can overwhelm the generation of Fourier-transform-limit pulses as the harmonic up-conversion number increases. Besides, for the generation of full coherent soft x-ray FEL pulses with harmonic up-conversion number of 270 corresponding to 1 nm, relative timing jitter between tens of fs-scale electron beam and external seed laser is the key issue in the feasibility study of the cascaded EEHG-HGHG scheme.

In this paper, we design and optimize the EEHG-HGHG configuration based on intensive start-to-end (S2E) numerical simulations. The effects of the relative timing jitter is discussed. This paper is organized as follows. In Sec.~\ref{sec:level2}, the baseline design of the soft x-ray beamline based on the cascaded EEHG-HGHG scheme is given. In addition, two cases of S2E electron beams with double-horn and fat-top current profiles are presented, respectively. Subsequently, intensive S2E FEL simulations with various current profiles and the sensitivity analysis of timing jitter are discussed in Sec.~\ref{sec:level3} and Sec.~\ref{sec:level4}, respectively. Finally, conclusions and prospects are given in Sec.~\ref{sec:level5}.

\begin{figure}
	\includegraphics[width=1\textwidth]{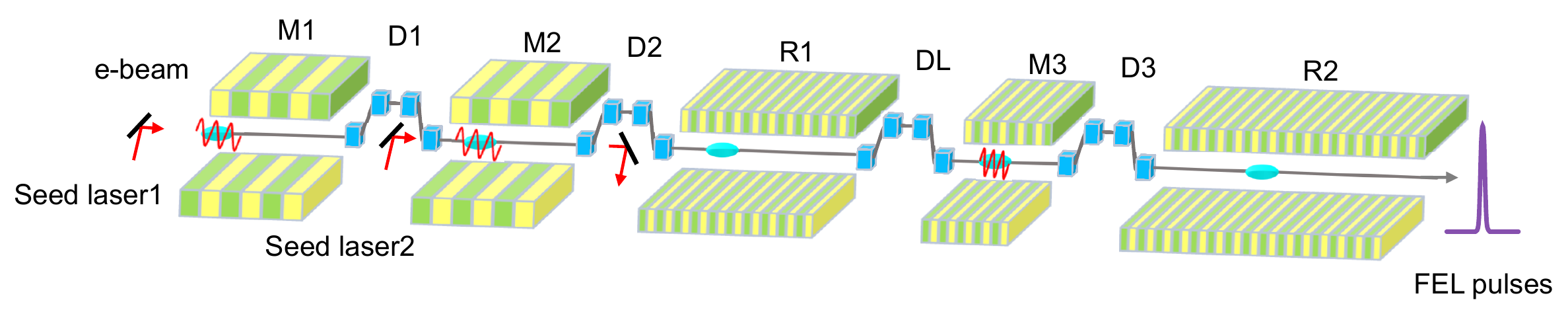}
	\caption{\label{fig:wide}Schematic layout of the cascaded EEHG-HGHG scheme for FEL-II undulator line at SHINE. (M: modulator, D: dispersion section, DL: delay line section, R: radiator)}
\end{figure}
\section{\label{sec:level2}The cascaded EEHG-HGHG scheme for SHINE}
Shanghai High-Repetition-Rate XFEL and Extreme Light Facility (SHINE) is a MHz-level XFEL facility based on the superconducting radiofrequency linac {\cite{Zhao:FEL2017}}. The electron bunches have a bunch charge of 100 pC, beam energy of 8 GeV, slice energy spread of 0.01\%, normalized emittance of 0.2 mm·mrad, and can be compressed to the peak current of 1.5 kA, {\cite{Zhou2019}}. Besides, SHINE consists of three beamlines FEL-I, FEL-II, and FEL-III, covering the photon energy of 3-15 keV, 0.4-3 keV, and 10-25 keV, respectively. The SASE and self-seeding operation modes are both adopted in FEL-I and FEL-III for generating hard x-ray FEL pulses. To meet the demands of users, FEL-II is designed as a soft x-ray beamline utilizing SASE and external seeding schemes. The cascaded EEHG-HGHG scheme is one of the promising candidates for FEL-II. The schematic layout of FEL-II is illustrated in Fig.~\ref{fig:wide}.

In the design and optimization of the EEHG-HGHG scheme, following the notation of Ref.{\cite{Xiang2009}}, $A_{1,2}=\Delta E_{1,2} / \sigma_{E}$ and $B_{1,2}=R_{56}^{(1,2)} k_{1,2} \sigma_{E} / E_{0}$ are the dimensionless parameters for tuning the maximum bunching factor, where $\Delta E_{1,2}$ are the energy amplitudes modulated by external seed lasers in M1 and M2, $\sigma_{E}$ is the initial slice energy spread, $E_{0}$ is the mean beam energy, $R_{56}^{(1,2)}$ are dispersion strengths for D1 and D2, $k_{1,2}$ are the wavenumbers of two seed lasers, and $K$ represents the ratio of the frequencies. Therefore, based on the basic principle of the EEHG scheme, the bunching factor of $a$th harmonic of seed laser can be written as
\begin{equation}
	\begin{aligned}
		b_{n, m}=& \mid e^{-(1 / 2)(n B_{1}+a B_{2})^{2}} J_{m}(-a A_{2} B_{2}) \\
		& \times J_{n}\left[-A_{1}\left(n B_{1}+a B_{2}\right)\right] \mid,
	\end{aligned}
\label{eq:1}
\end{equation}
where $a=n+mK$, $n$ is a negative integer, and $m$ is a positive integer. According to the optimization method in Refs.{\cite{Stupakov2009,Xiang2009}}, the maximum bunching factor can be achieved when
\begin{equation}
	B_{2}=\frac{m+0.81 m^{1 / 3}}{a A_{2}},
\label{eq:2}
\end{equation}
\begin{equation}
	A_1\left[J_{n-1}\left(A_1\ \xi\right)-J_{n+1}\left(A_1\ \xi\right)\right]=2\xi J_n\left(A_1\xi\right).
\label{eq:3}
\end{equation}

In addition, the optimal relationship of the first and the second dispersion strengths should be derived as
\begin{equation}
	B_{2}=\frac{n}{a} B_{1}-\frac{\xi}{a},
	\label{eq:4}
\end{equation}
where $\xi$ is the optimized solution of Eq.~(\ref{eq:3}) As the energy modulation amplitude $A_{1}$ and harmonic number $a$ increases, $\xi$ can significantly decrease, thus the Eq.~(\ref{eq:4}) can approximately rewrite as $B_2\approx(\mid n\mid/a){\ B}_1$. For a specific harmonic number $a$, the maximum bunching factor can be obtained for $n=-1$, when both Eq.~(\ref{eq:2}) and Eq.~(\ref{eq:3}) are satisfied. Further, according to the optimization process, bunching is more sensitive to $B_{1}$, which is a function of $n$ and decreases as $\mid n\mid$ increases, and smaller $B_{1}$ is easier to implement in engineering practice. At the same time, bunching decreases as $\mid n\mid$ increases, so $A_{1}$ should be large enough to ensure sufficient bunching of specific harmonic.

\begin{table}[!hbt]
    \centering
	\caption{\label{tab:table1}Main electron beam, seed laser, and undulator line parameters of FEL-II design.}
	\begin{tabular}{ccc}
        \toprule
		Parameters				&Value	&Unit\\
		\midrule
		{\textbf{Electron beam}}&		&\\
		Energy					&8		&GeV\\
		Slice energy spread 	&0.01	&\%\\
		Normalized emittance  	&0.2	&mm mrad\\
		Bunch charge			&100	&pC\\
		Peak current (Gaussian) &1.5	&kA\\
		Bunch length (FWHM)			&60		&fs\\
		{\textbf{Seed laser}}	&		&\\
		Seed wavelength			&270	&nm\\
		Rayleigh length			&3.52	&m\\
		Seed power				&10s	&GW\\
		Duration (FWHM)		&20		&fs\\
		{\textbf{Undulator line}}&		&\\
		M1 and M2 period		&0.24	&m\\
		M3, R1, and R2 period	&0.068	&m\\
		Dispersion strength of D1				&1.4	&mm\\
		Dispersion strength of D2				&80		&$\mu$m\\
		Dispersion strength of D3				&2		&$\mu$m\\
		Dispersion strength of DL				&14		&$\mu$m\\
		\bottomrule
	\end{tabular}
\end{table}

In the first stage of EEHG, two UV seed lasers of 10 GW-level modulate the electron beam in two modulators with a period of 0.24 m. The external seed lasers can produce energy modulation amplitude of $A_{1}$ = 9 and $A_{2}$ = 6, respectively. After passing dispersion sections of optimal values, the electron beam can generate 5 nm fully coherent x-rays with harmonic up-conversion number of 54. Considering practical limitations of the length of dispersion sections, and the IBS effect decreasing the bunching factor, the first stage EEHG is operated in high-order mode {\cite{Zhou2016}}, corresponding to $n=-3$ for Eq.~(\ref{eq:4}). Meanwhile, the second stage HGHG produces full coherent FEL pulses of 1 nm with a harmonic up-conversion number of 5, corresponding to the energy modulation amplitude of about 5. Since the first stage EEHG is considered as the seed laser, its FEL output is not saturated in the order of 10 GW. With the “fresh bunch” technique {\cite{Yu1997}}, the full coherent soft x-ray is generated in the second stage HGHG. Overall, the main electron beam, seed laser, and undulator line parameters of FEL-II are listed in Table~\ref{tab:table1}.

\begin{figure}[htbp]
    \centering
	\includegraphics[width=0.8\textwidth]{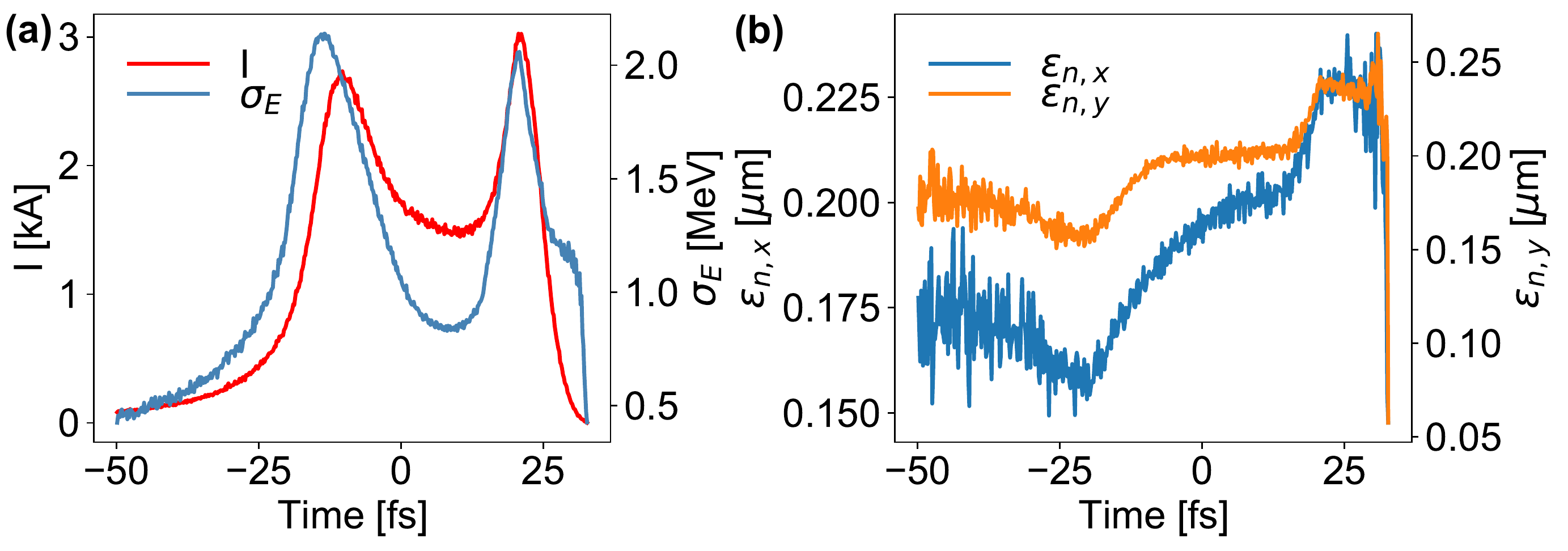}
	\caption{\label{fig:double} The current, energy spread, and emittance of the start-to-end electron beam in the numerical simulations. The electron beam has a double-horn current profile with its head on the right.}
\end{figure}

\begin{figure}[htbp]
    \centering
	\includegraphics[width=0.8\textwidth]{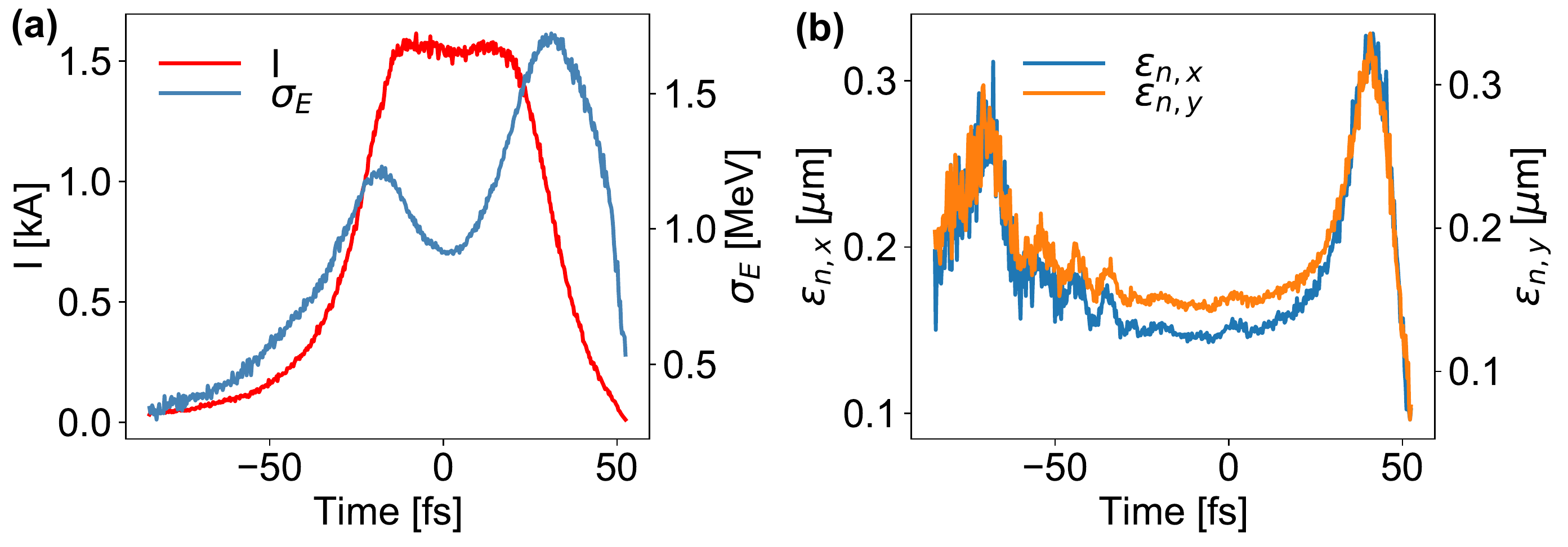}
	\caption{\label{fig:flat} The current, energy spread, and emittance of the start-to-end electron beam in the numerical simulations. The electron beam has a flat-top current profile with its head on the right.}
\end{figure}

Start-to-end particle tracking has been employed to study the electron with various practical effects. ASTRA {\cite{flottmann2011astra}} is used to track electron beams for the injector, and then ELEGANT {\cite{Borland2000}} is utilized for the main linac and bunch compressors until the entrance of the undulator line. One S2E electron beam current profile case has two spikes, i.e., a double-horn profile and its longitudinal phase space properties are shown in Fig.~\ref{fig:double}. The energy distribution, in this case, is almost identical to the double-horn profile, with a large energy spread at the peak position. For the FEL community, a more linearized longitudinal phase space of the electron beam is more favorable for generating stable FEL pulses, especially for the seeding scheme. A longitudinal phase space manipulation for inhibiting current-spike formation is proposed for generating electron beams with a flat-top distribution {\cite{ZHU2022}} as another case, i.e., a fat-top profile, whose beam properties are shown in Fig.~\ref{fig:flat}.

\section{\label{sec:level3}Three-dimensional S2E simulations with relative timing jitter}
In the design and optimization of an externally seeded FEL, the beam-to-beam jitter from the linac, spatial and temporal jitter of the seed laser, and electron beam should keep within a reasonable range. Their effects on the FEL performance are accurately calculated to evaluate the feasibility of the cascaded EEHG-HGHG scheme. The parameters of the electron beam, such as emittance, beam current, slice energy spread, etc., on the femtosecond scale of the laser-beam interaction limit the FEL performance with the timing jitter budget {\cite{Lechner2014}}. Therefore, numerical S2E simulations of three-dimensional (3D) effects are performed by GENESIS {\cite{REICHE1999}} using two types of S2E electron beams, i.e., those with double-horn and flat-top current profiles as shown in Fig.~\ref{fig:double} and Fig.~\ref{fig:flat}, respectively. Note that the nominal case with ideal Gaussian current profile is considered as a comparison. The projected emittance and energy spread are set the same to meet the design specifications of the linac, as listed in Table~\ref{tab:table1}. The double-horn and flat-top current profiles are compared to demonstrate the stability of the output FEL performance in the first stage EEHG and the second stage HGHG.

\begin{figure}
    \centering
	\includegraphics[width=0.22\textwidth]{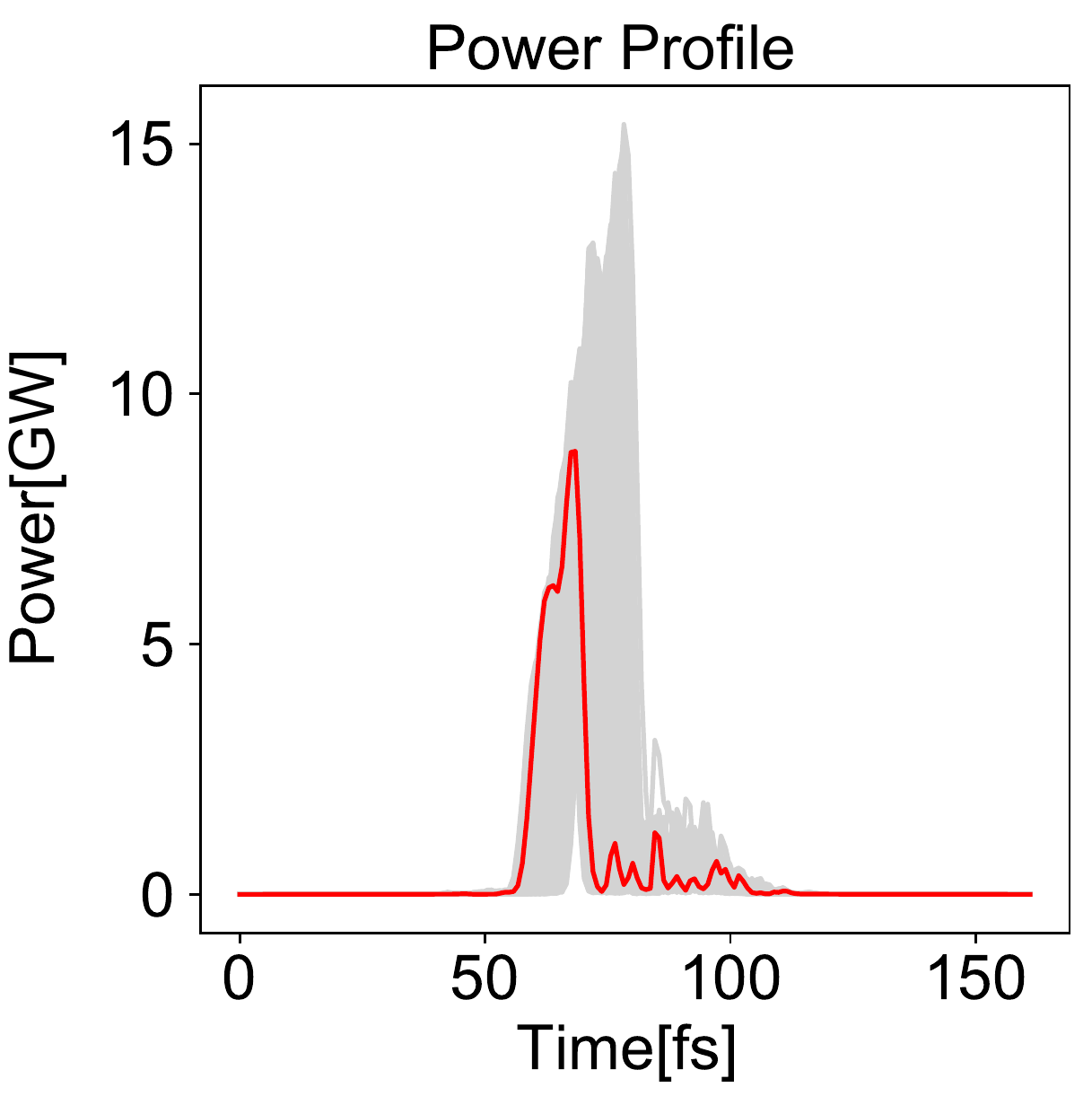}
	\includegraphics[width=0.22\textwidth]{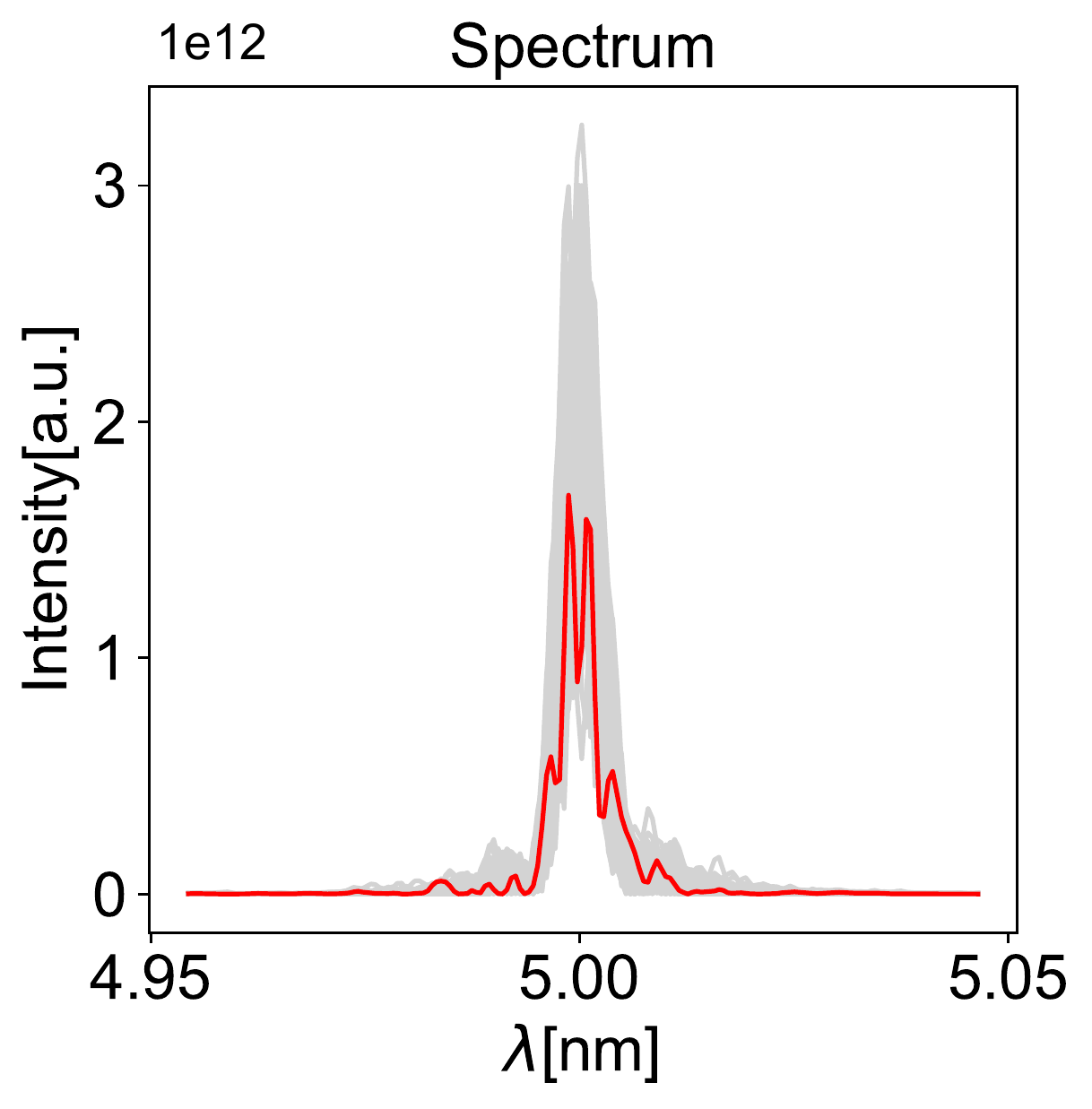}\\
	\hspace*{-10pt}
	\includegraphics[width=0.22\textwidth]{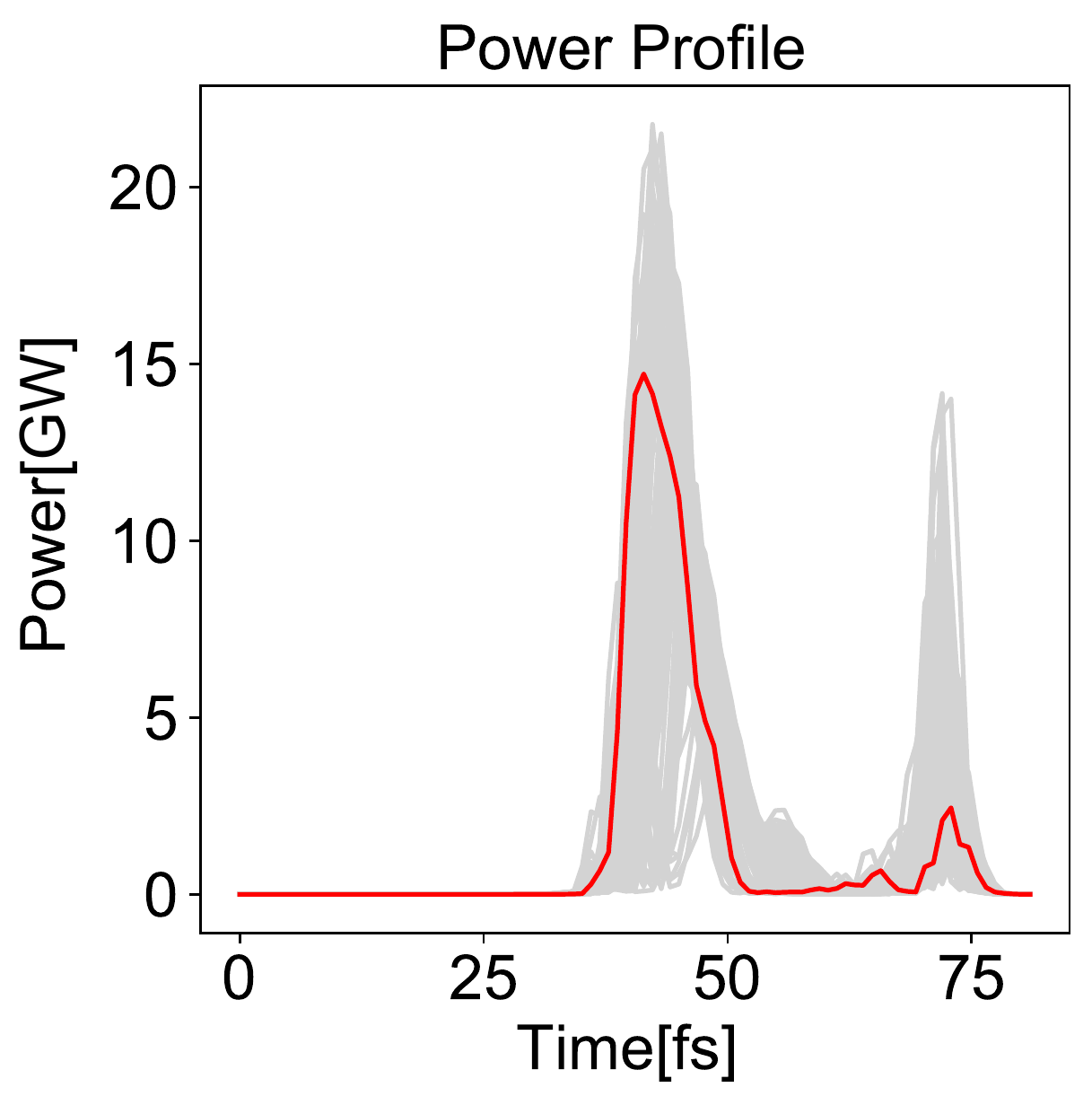}
	\includegraphics[width=0.21\textwidth]{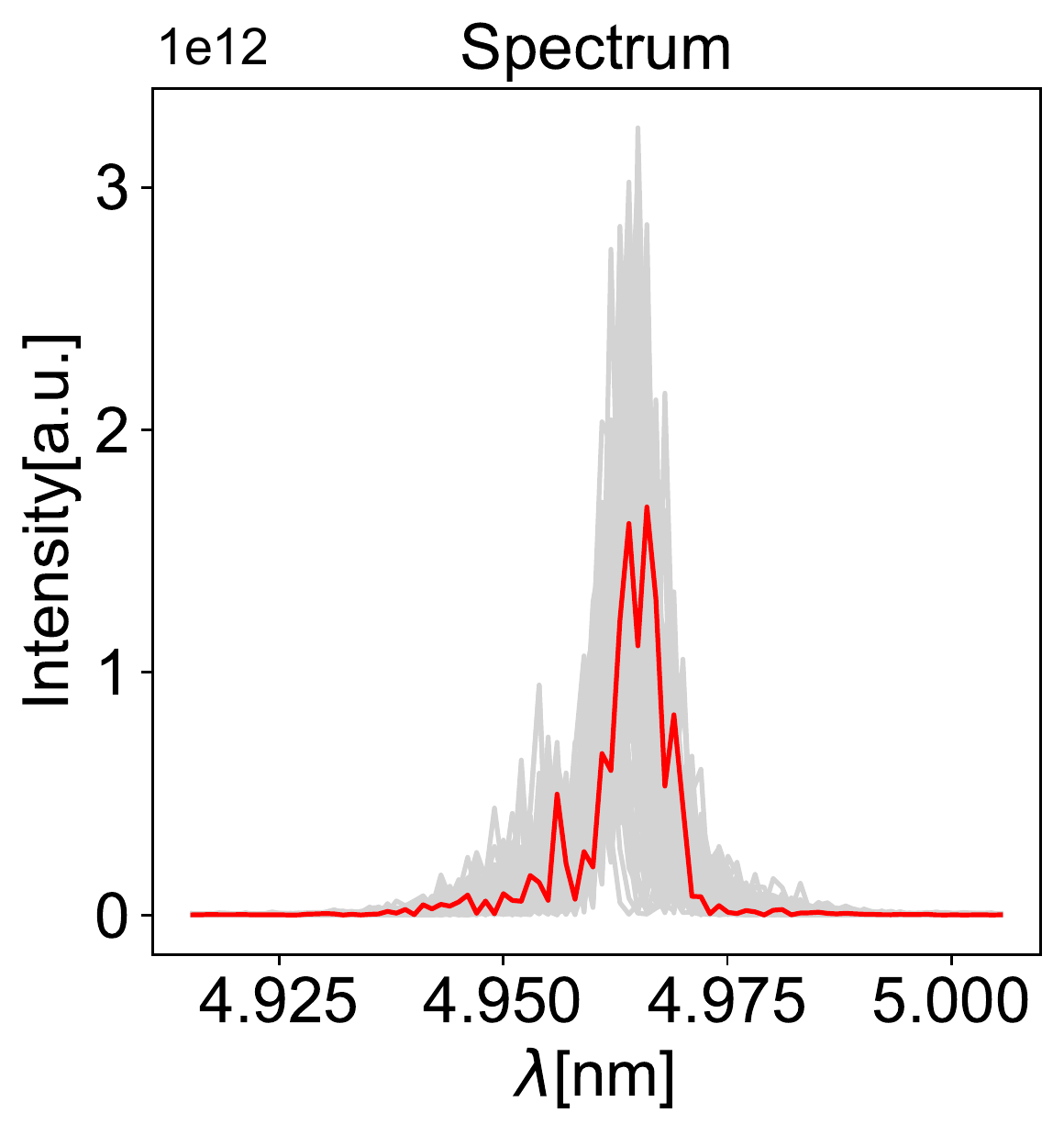}\\
	\includegraphics[width=0.21\textwidth]{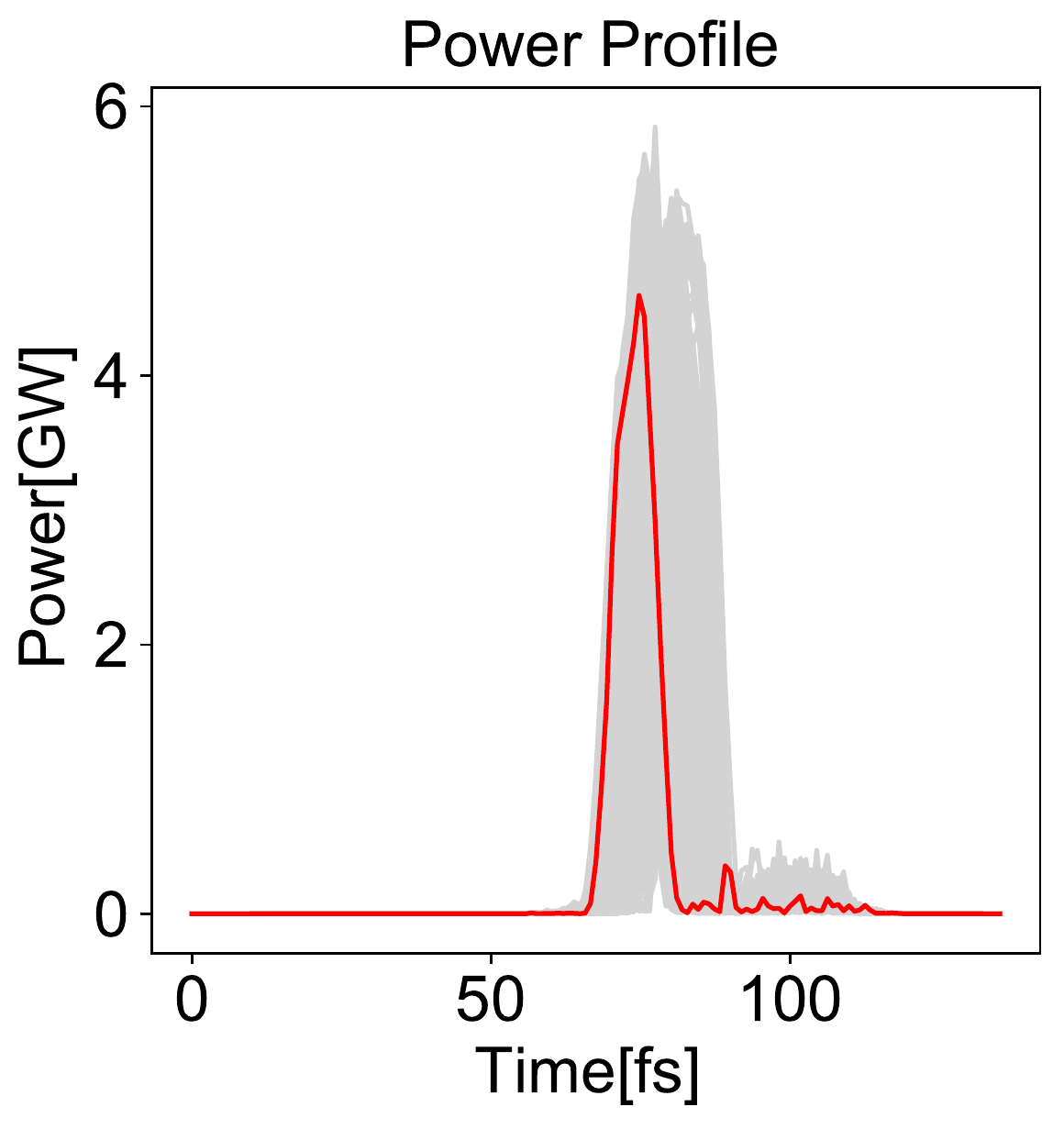}
	\includegraphics[width=0.22\textwidth]{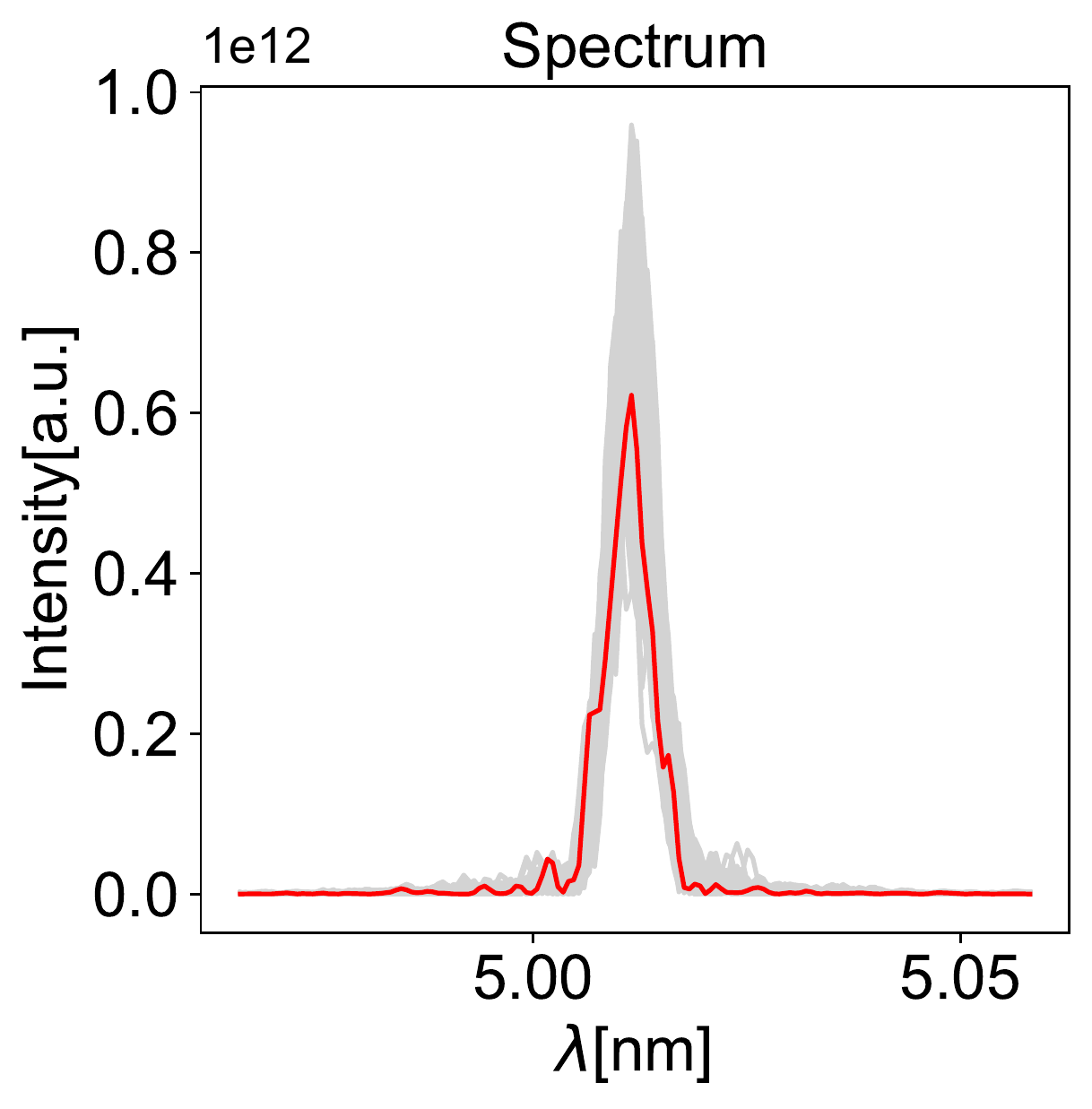}
	\caption{\label{fig:EEHG}The 100 shots numerical simulations of the first stage EEHG FEL performance with Gaussian current profile (the top) and 100 shots start-to-end simulations with double-horn (the middle) and flat-top (the bottom) current profiles, respectively.}
\end{figure}

Two seed lasers of 270 nm modulate the same electron beam within the seed laser wavelength, leading to sinusoidal energy modulation. In general, two UV seed lasers can be split precisely from optics, the relative timing jitter is neglectable. For the sake of simplicity, there is no arrival timing jitter of each electron beam. Meanwhile, the relative timing jitter of the seed lasers and electron beam at the entrance of modulators only is considered. Due to the S2E electron beam properties with different energy distributions or current profiles, the seed lasers with peak power in the order of 10 GW are utilized for generating energy modulation amplitude of $A_{1}$ = 9 and $A_{2}$ = 6, respectively, as mentioned in Sec.~\ref{sec:level2}. The pulse duration of the seed laser is chosen to be 20 fs (FWHM) because the bunch lengths of different profiles are around 60 fs (FWHM), and the electron beam should delay by 20 fs maintaining the radiation pulses with “fresh bunch” in the cascaded EEHG-HGHG scheme. To investigate the relative timing jitter effects on FEL performance, a Gaussian random distribution of timing jitter of 3 fs (RMS) which is likely achieved in a state-of-the-art seed laser system, is added to seed lasers at the entrance of the modulators. In addition, all parameters used for S2E numerical simulations are accurately tuned and optimized to obtain the optimal FEL pulses, and these slightly different parameters are considered as ideal configurations for different current profiles without relative timing jitter.

\subsection{The first stage EEHG}
The FEL performance of the first stage EEHG simulated by GENESIS with Gaussian, double-horn, and flat-top profiles are shown in Fig.~\ref{fig:EEHG}, respectively. Figure ~\ref{fig:EEHG} shows that FEL spectra have shift of the central wavelength in both double-horn and flat-top current profiles owing to different electron beam parameters, seed laser, and optimal parameters of undulator lines {\cite{WANG201456,Paraskaki2019}}. Due to the slippage effect of the first stage and the delay section, two spikes are evident in the power profile with the double-horn current distribution corresponding to the peak current of around 3 kA, where the EEHG FEL radiation is unable to suppress the SASE from the unseeded part {\cite{Hemsing2019}}. Thus, the relative timing jitter can directly lead to this undesirable situation. Besides, the statistical S2E simulated results of the first stage EEHG are presented in Table~\ref{tab:table2}. The mean pulse energy of 5nm FEL output with Gaussian profile, double-horn profile, and flat-top profile are 115.47 $\mu$J, 119.66 $\mu$J, and 46.17$\mu$J, corresponding to standard deviations of 20.75 $\mu$J, 27.34 $\mu$J, and 3.49 $\mu$J, or the fluctuations of 17.97\%, 22.84\%, and 7.56\%, respectively. Additionally, the mean pulse energy of both the Gaussian and double-horn profiles is significantly larger than that of the flat-top, since the first stage is to be used as a seed laser for the second stage not saturated as mentioned above. Therefore, the first stage EEHG is considered as an intermediate section in S2E numerical simulations. The pulse energy stability of the flat-top current profile is more stable, and its fluctuation is only 33\% of that of the double-horn profile.

\begin{table}
    \centering
	\caption{\label{tab:table2}The statistical numerical simulations of the output FEL pulse energy with Gaussian profile in the first and second stages and statistical start-to-end simulated results with double-horn profile and flat-top profile, respectively.}
	\begin{tabular}{ccccc}
        \toprule
		&Profile&Mean ($\mu$J)&SD ($\mu$J)&Fluctuation (\%)\\
		\midrule
		{\textbf{Stage1 EEHG}}&Gaussian&115.47&20.75&17.97\\
		&Double-horn&119.66&27.34&22.84\\
		&Flat-top&46.17&3.49&7.56\\
		{\textbf{Stage2 HGHG@20m}}&Gaussian&51.08&16.71&32.72\\
		&Double-horn&38.90&16.40&42.14\\
		&Flat-top&16.15&5.64&34.95\\
		{\textbf{Stage2 HGHG@30m}}&Gaussian&225.77&24.81&11.50\\
		&Double-horn&229.08&22.68&9.90\\
		&Flat-top&145.37&29.46&20.26\\
		\bottomrule
	\end{tabular}
\end{table}

\begin{figure}
    \centering
	\includegraphics[width=0.215\textwidth]{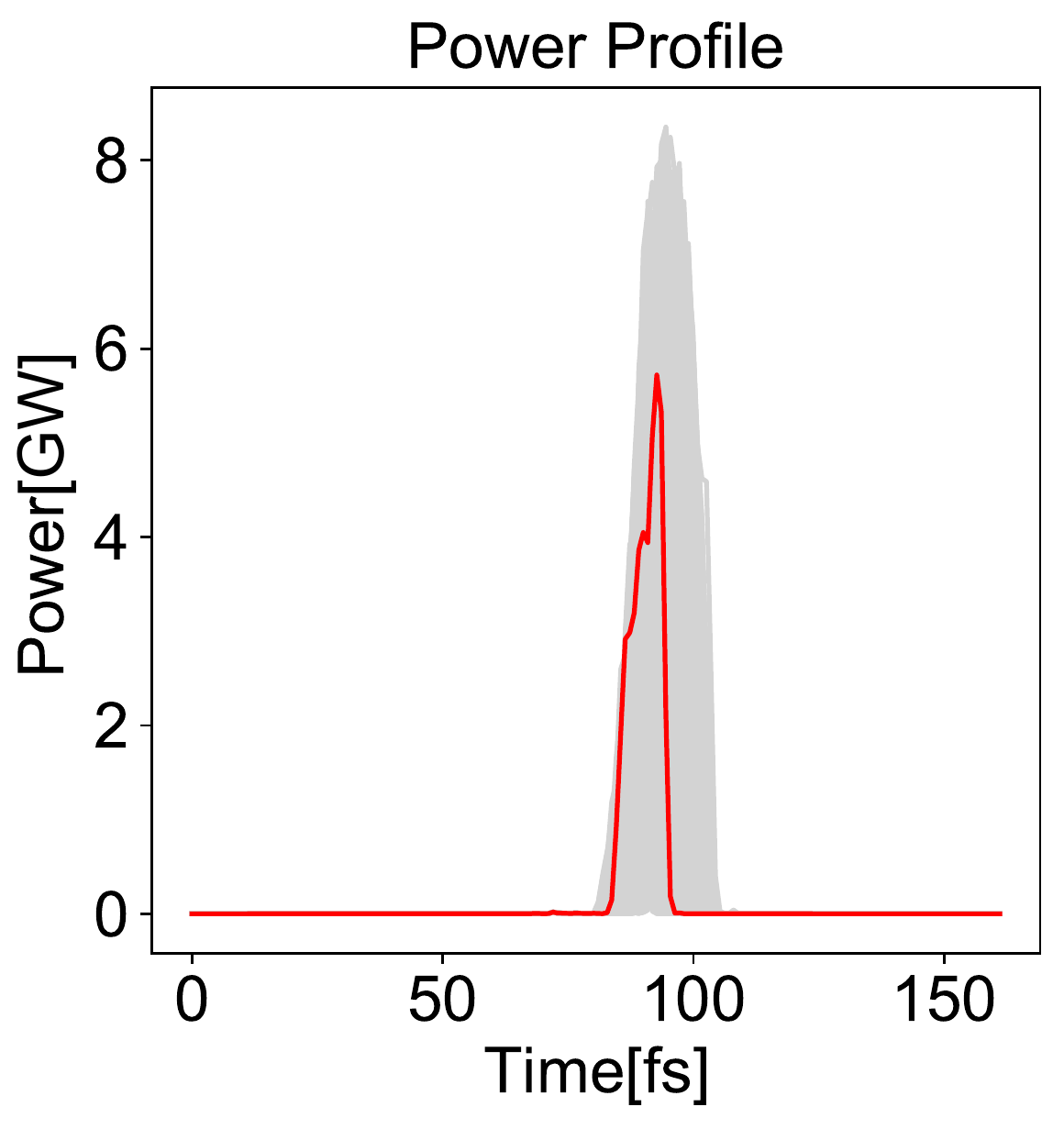}
	\includegraphics[width=0.24\textwidth]{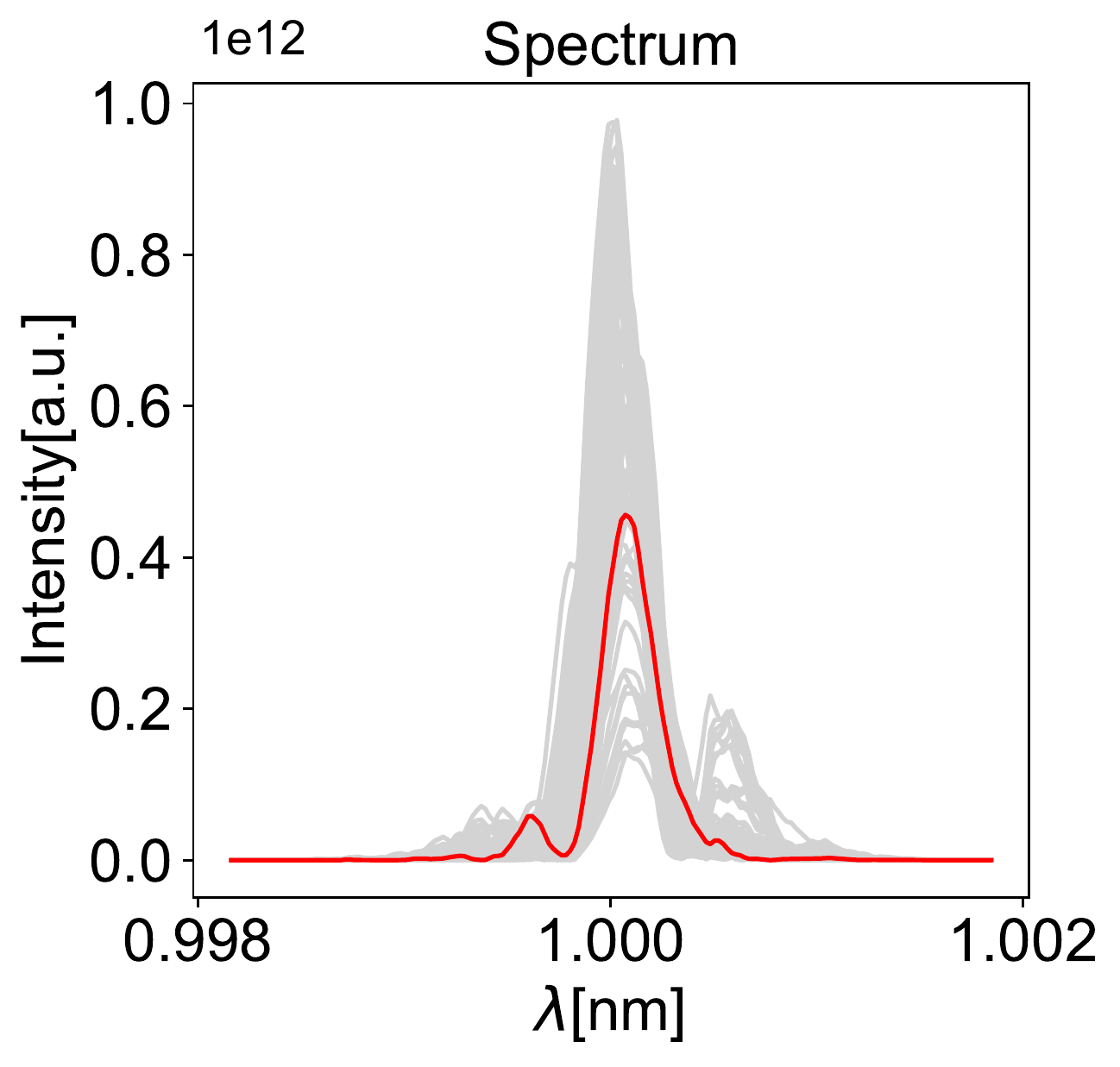}\\
	\hspace*{-15pt}
	\includegraphics[width=0.225\textwidth]{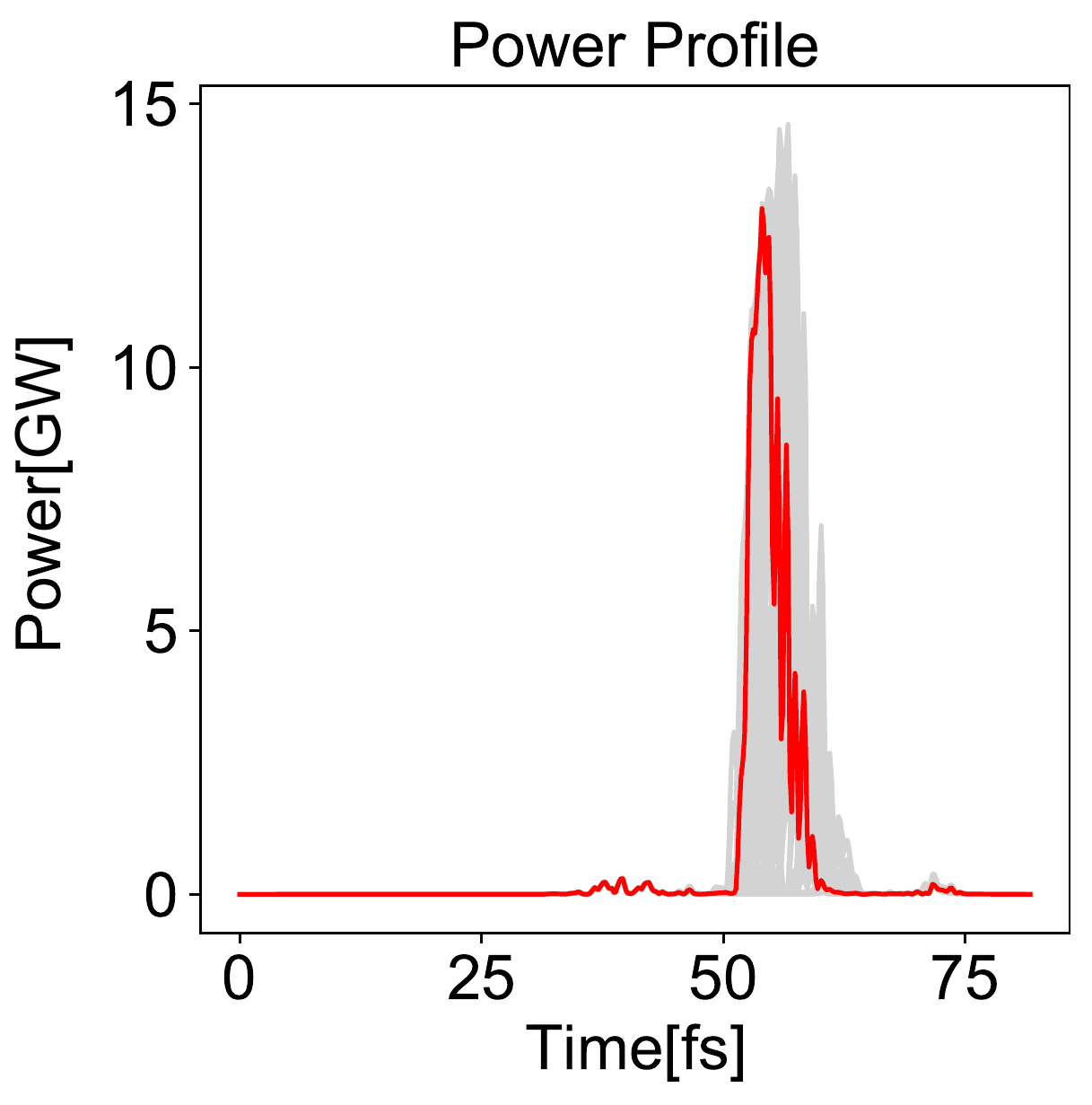}
	\includegraphics[width=0.22\textwidth]{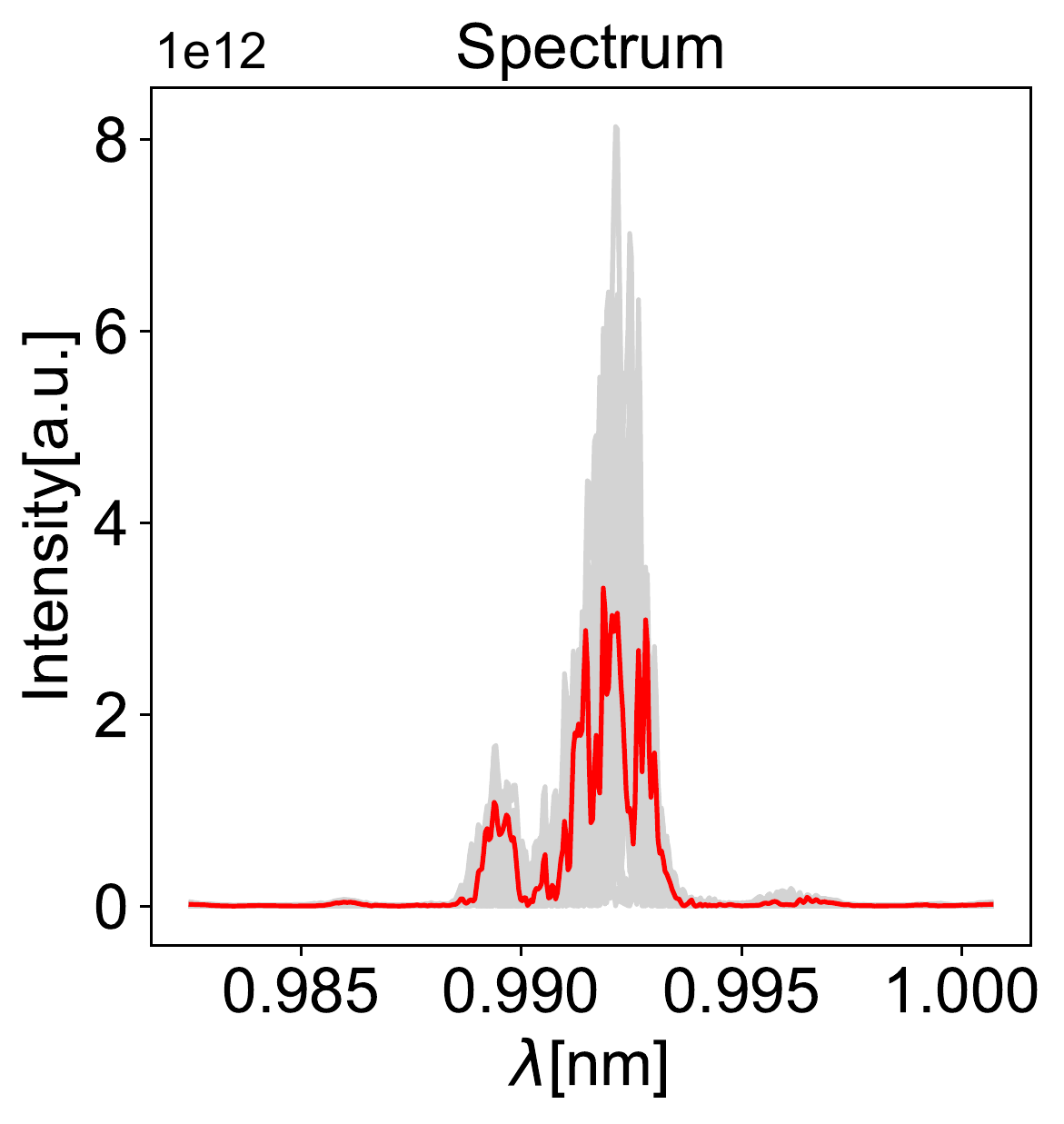}\\
	\includegraphics[width=0.216\textwidth]{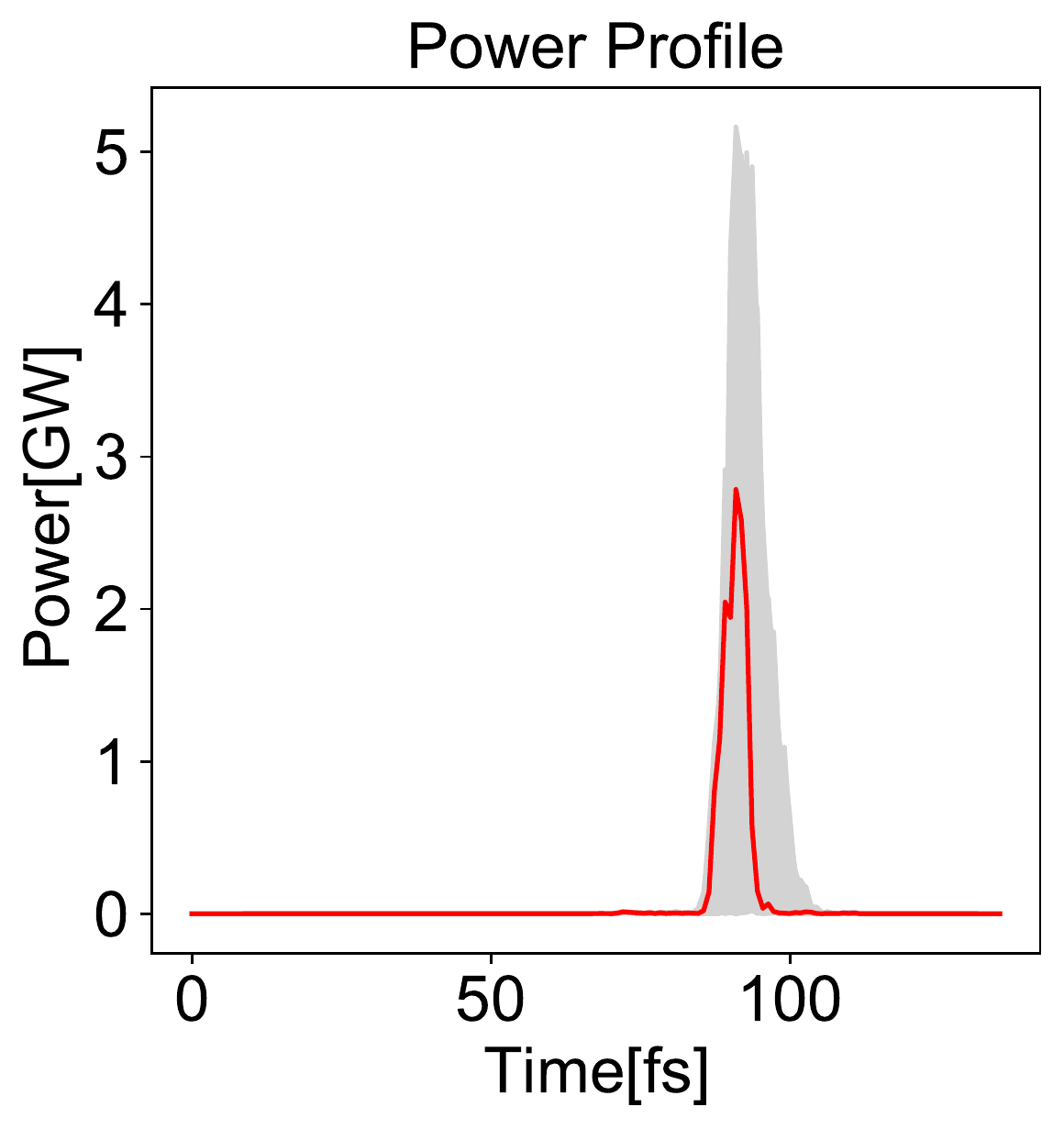}
	\includegraphics[width=0.233\textwidth]{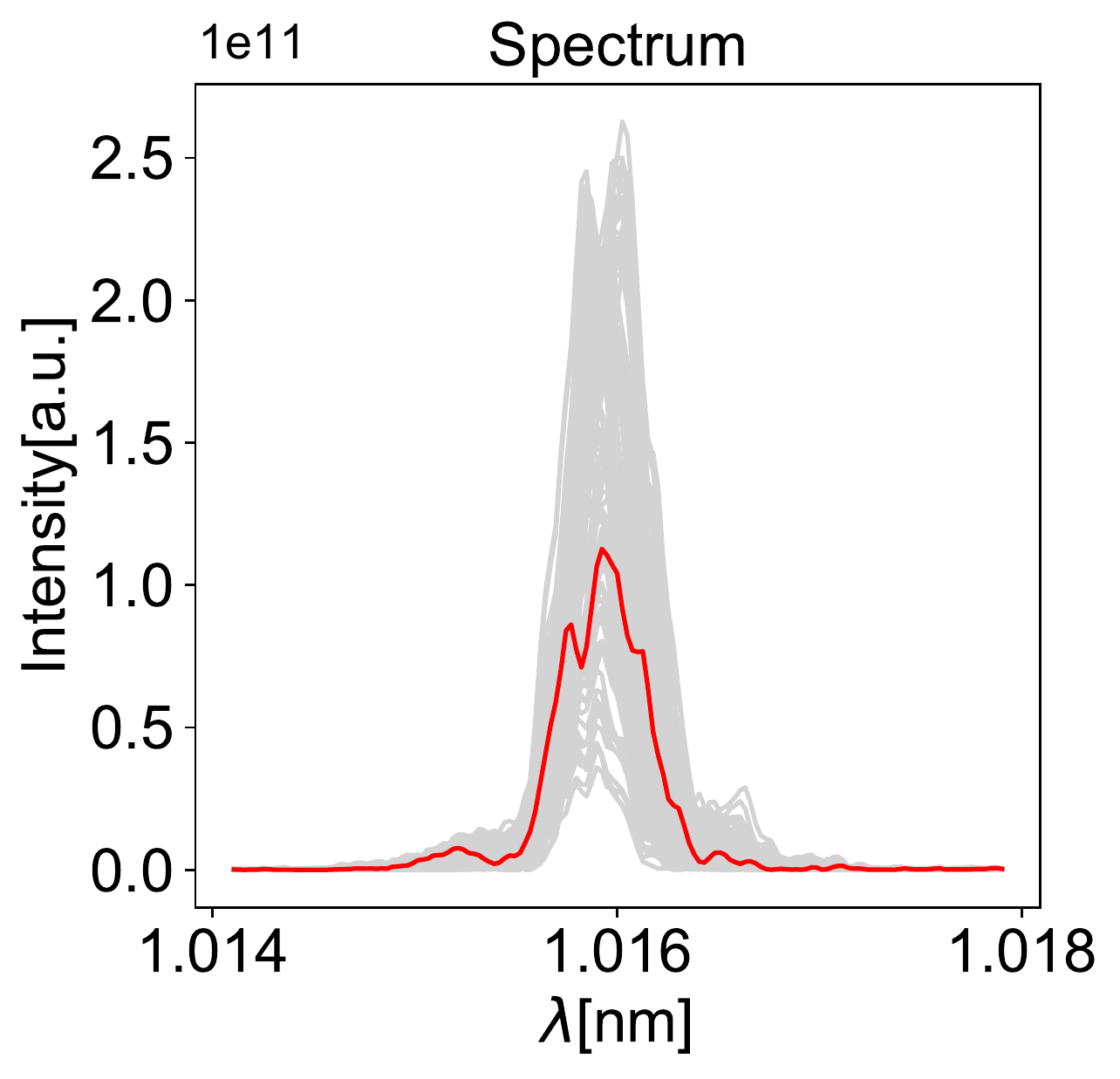}
	\caption{\label{fig:HGHG20}The 100 shots numerical simulations of the second stage HGHG FEL performance at 20 m along the radiators with Gaussian current profile (the top) and 100 shots start-to-end simulations with double-horn (the middle) and flat-top (the bottom) current profiles, respectively. Each gray line corresponds to each simulation, and the red line corresponds to one of the shots.}
\end{figure}

\begin{figure}
    \centering
	\includegraphics[width=0.28\textwidth]{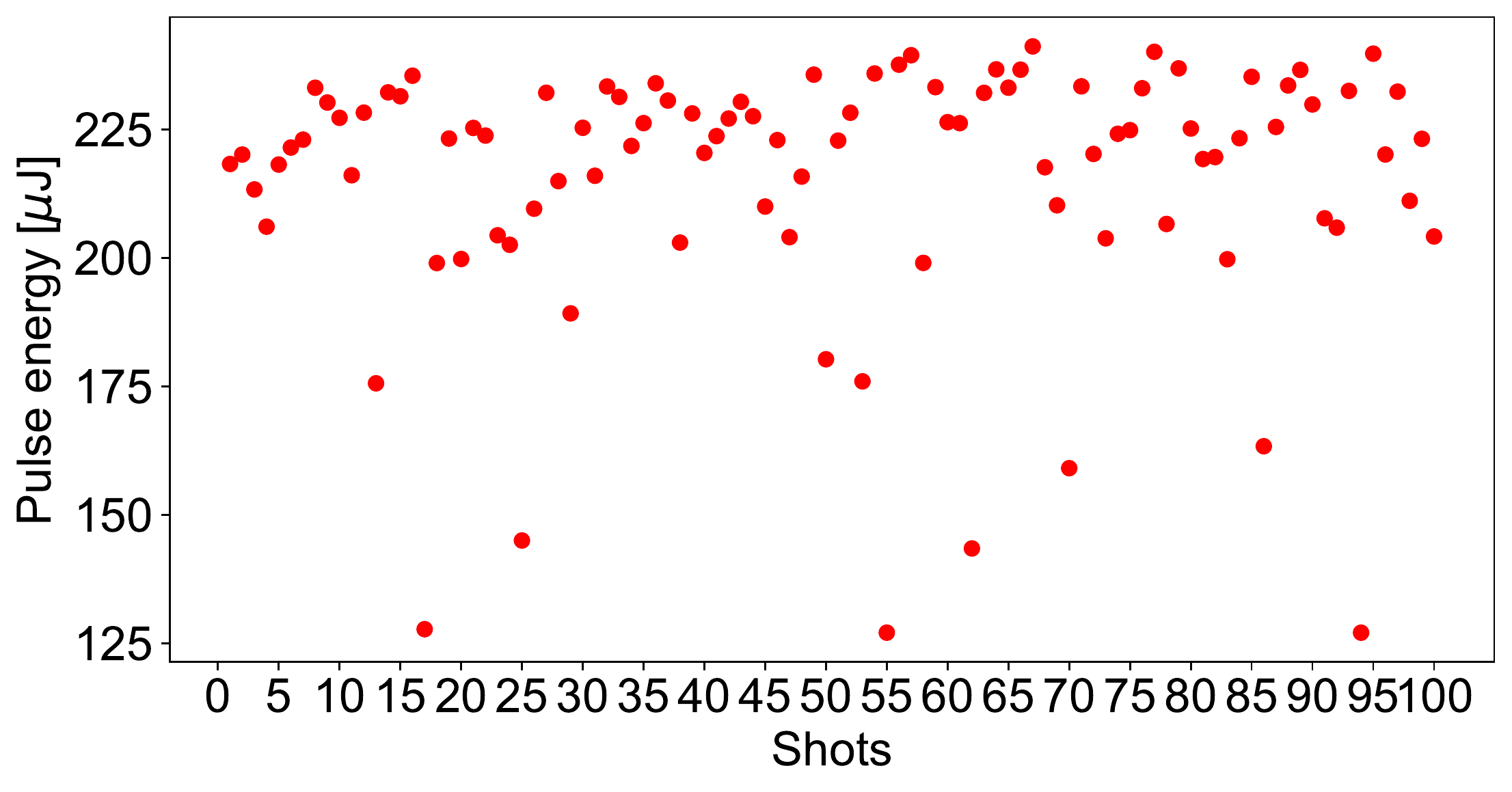}
	\includegraphics[width=0.16\textwidth]{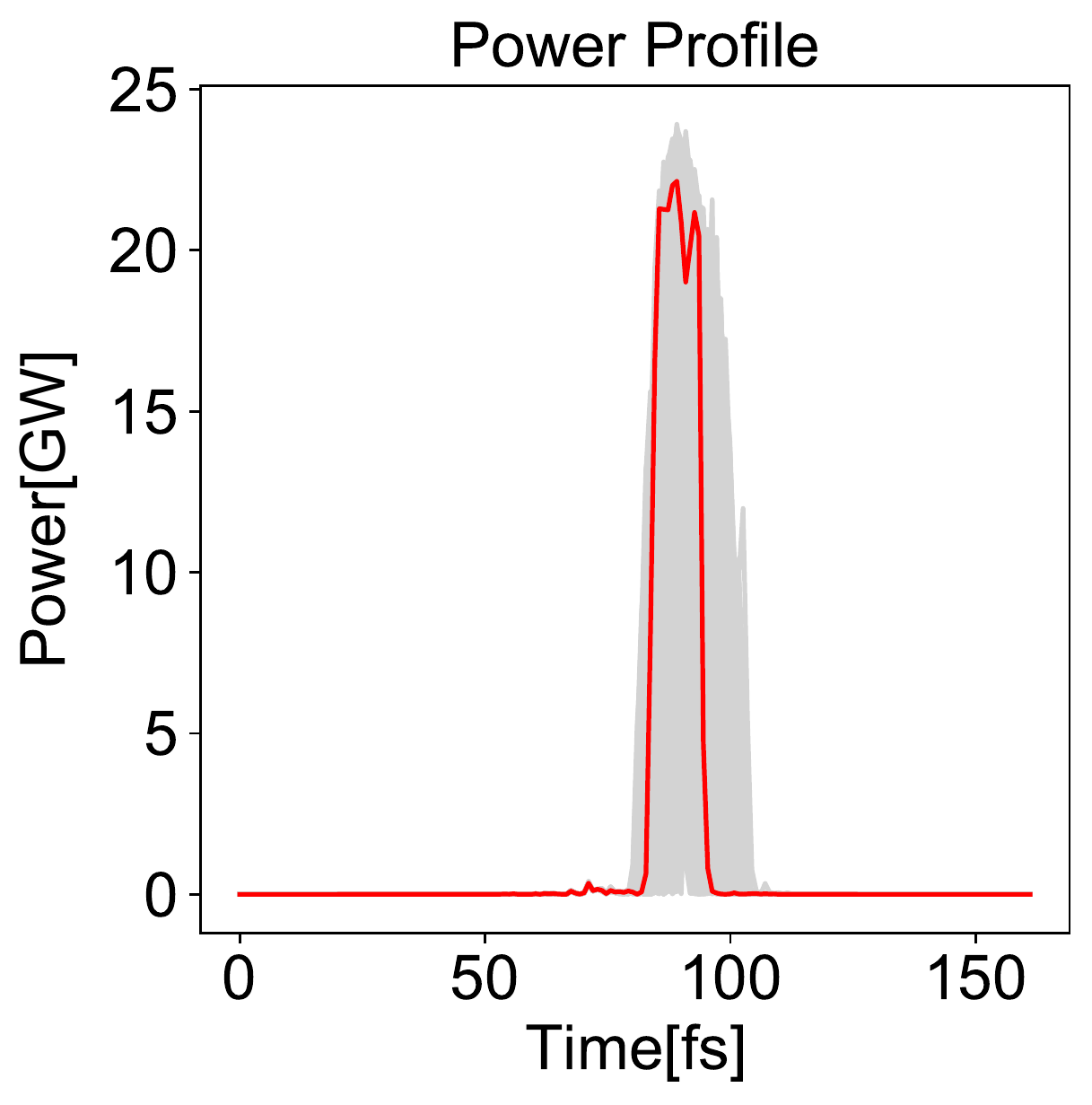}
	\quad
	\includegraphics[width=0.28\textwidth]{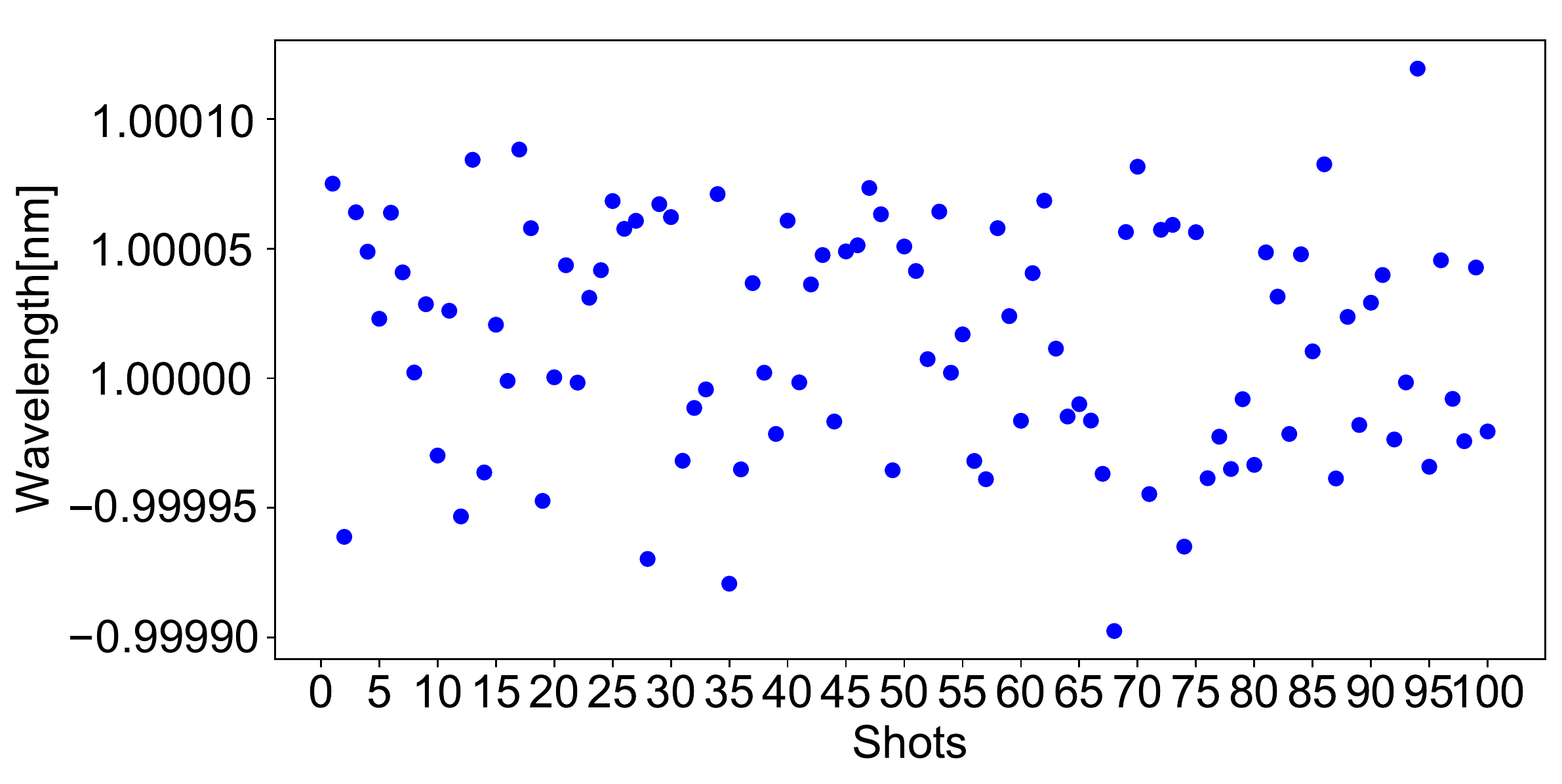}
	\includegraphics[width=0.16\textwidth]{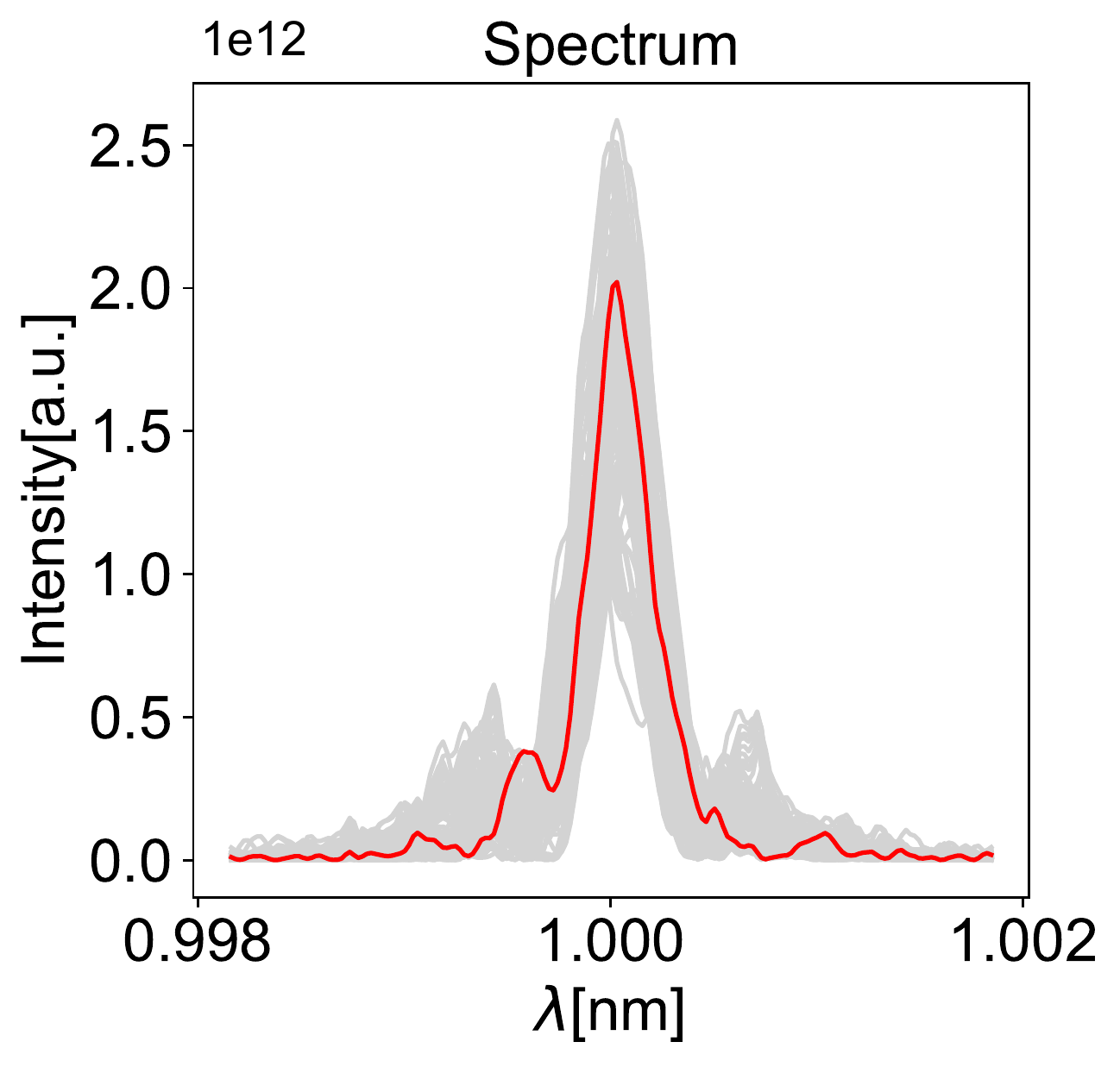}
	\caption{\label{fig:gauss1} The 100 shots numerical simulations of the second stage HGHG FEL performance at 30 m along the radiators with Gaussian current profile.}
\end{figure}

\begin{figure}
    \centering
	\includegraphics[width=0.28\textwidth]{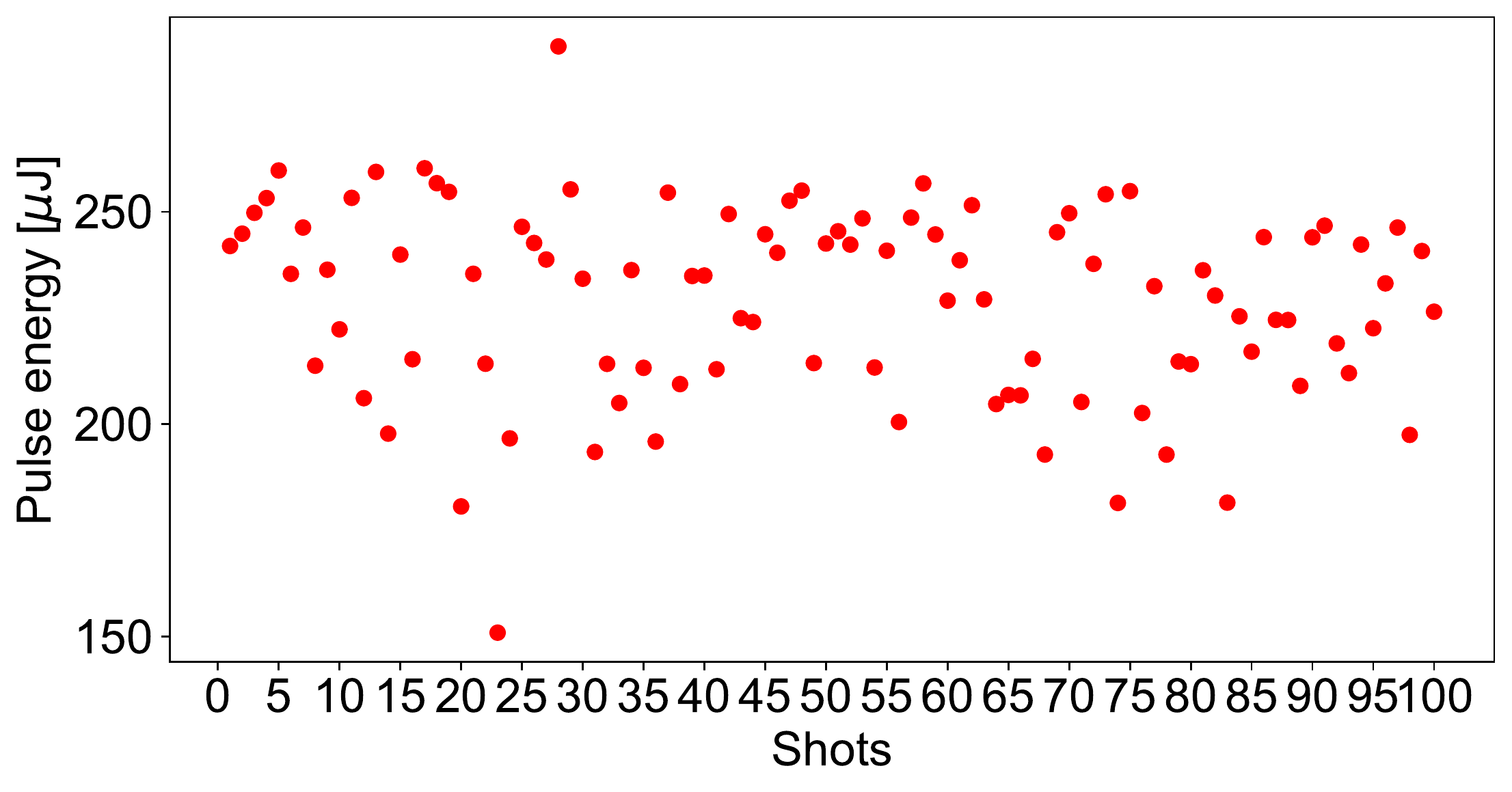}
	\includegraphics[width=0.16\textwidth]{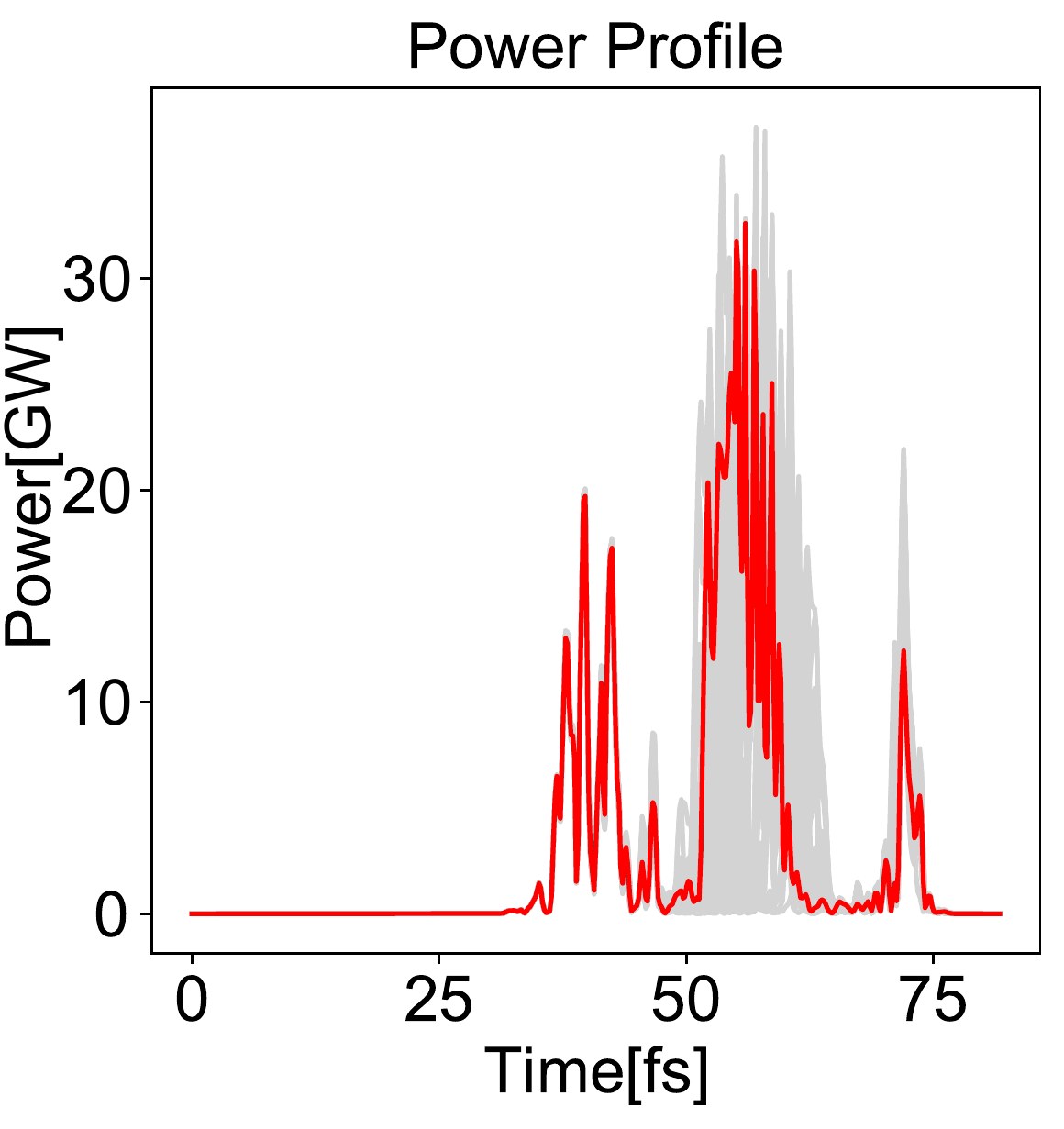}
	\quad
	\includegraphics[width=0.28\textwidth]{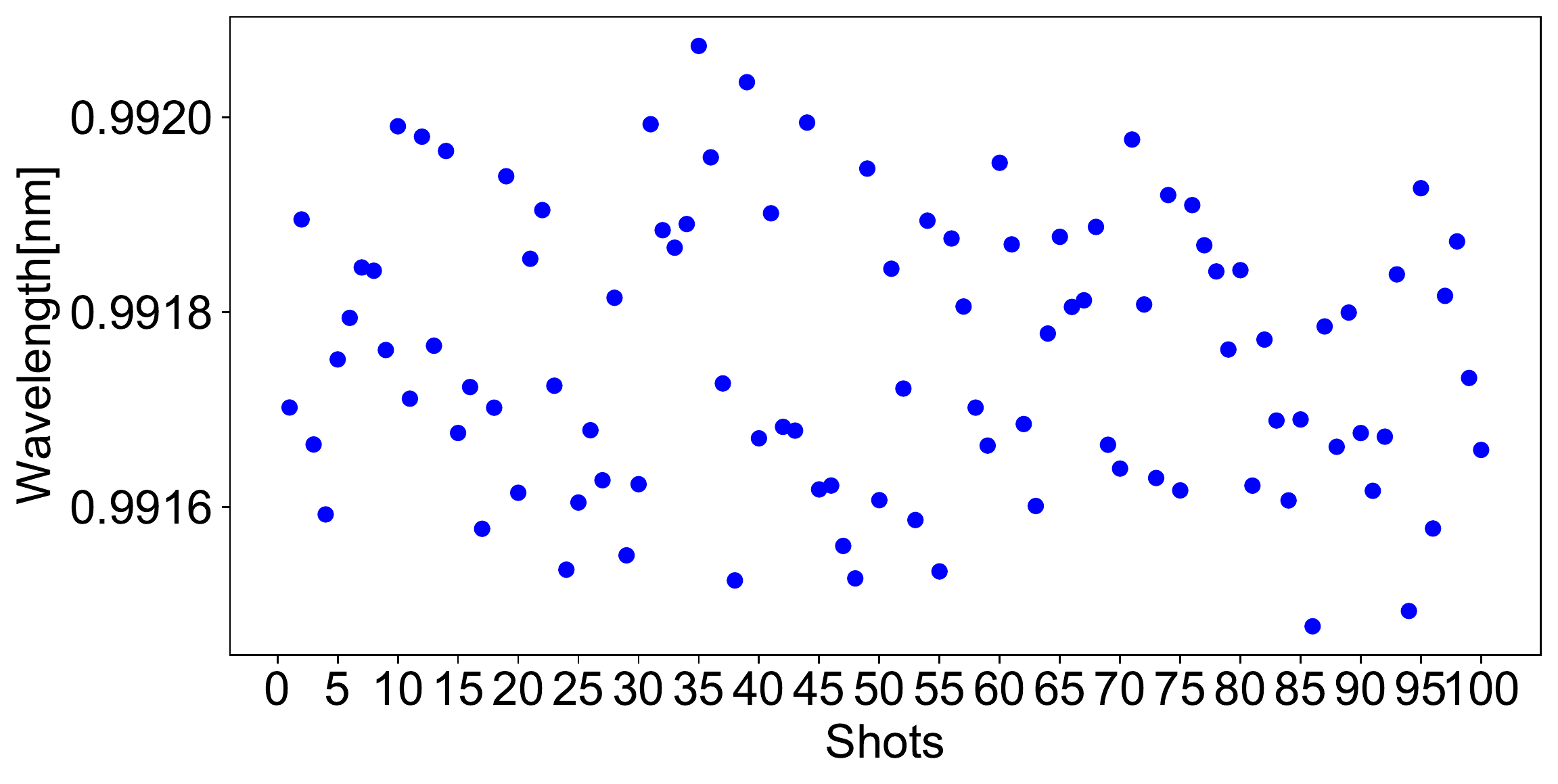}
	\includegraphics[width=0.16\textwidth]{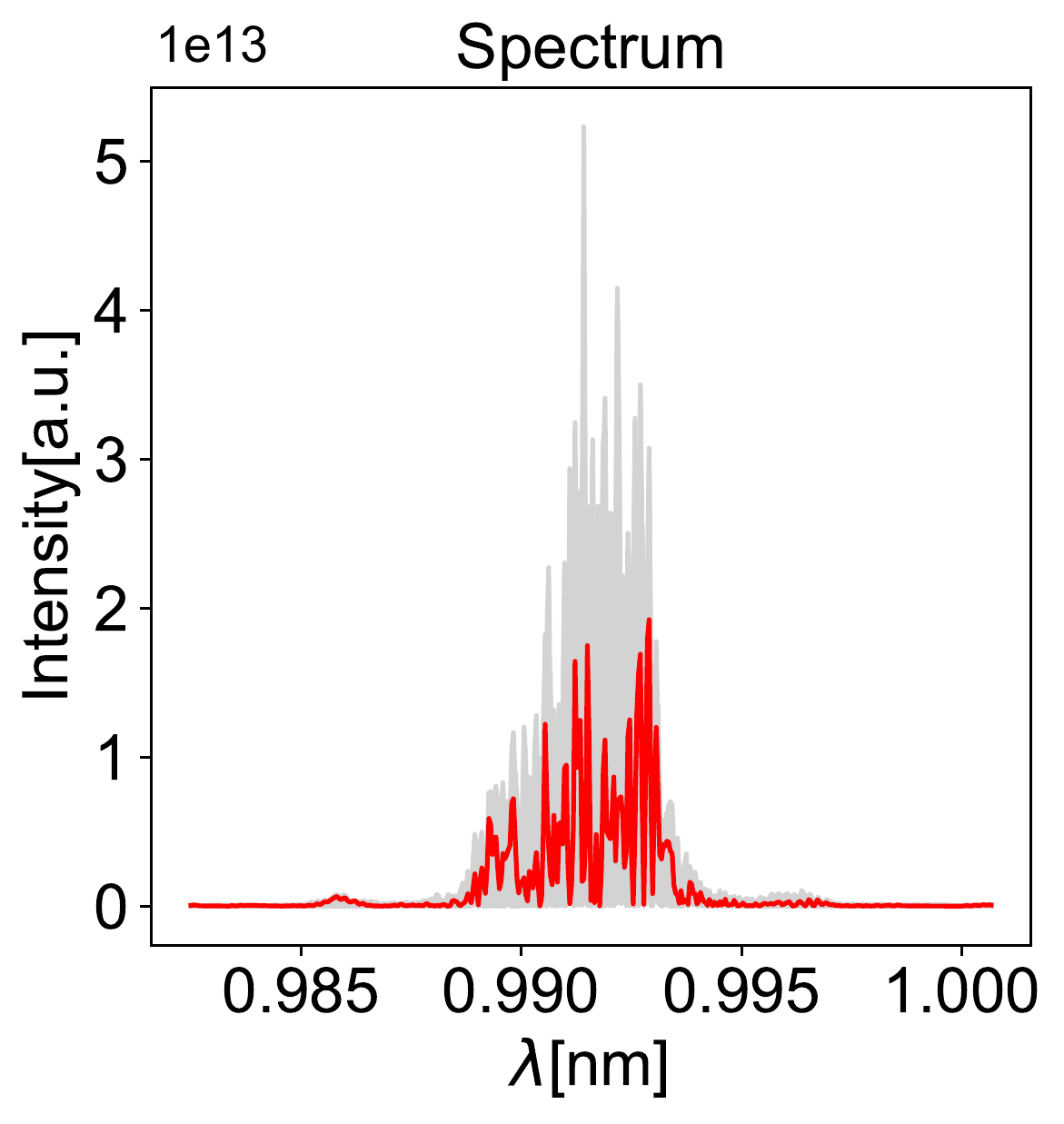}
	\caption{\label{fig:double1} The 100 shots start-to-end simulations of the second stage HGHG FEL performance at 30 m along the radiators with double-horn current profile.}
\end{figure}

\begin{figure}
    \centering
	\includegraphics[width=0.28\textwidth]{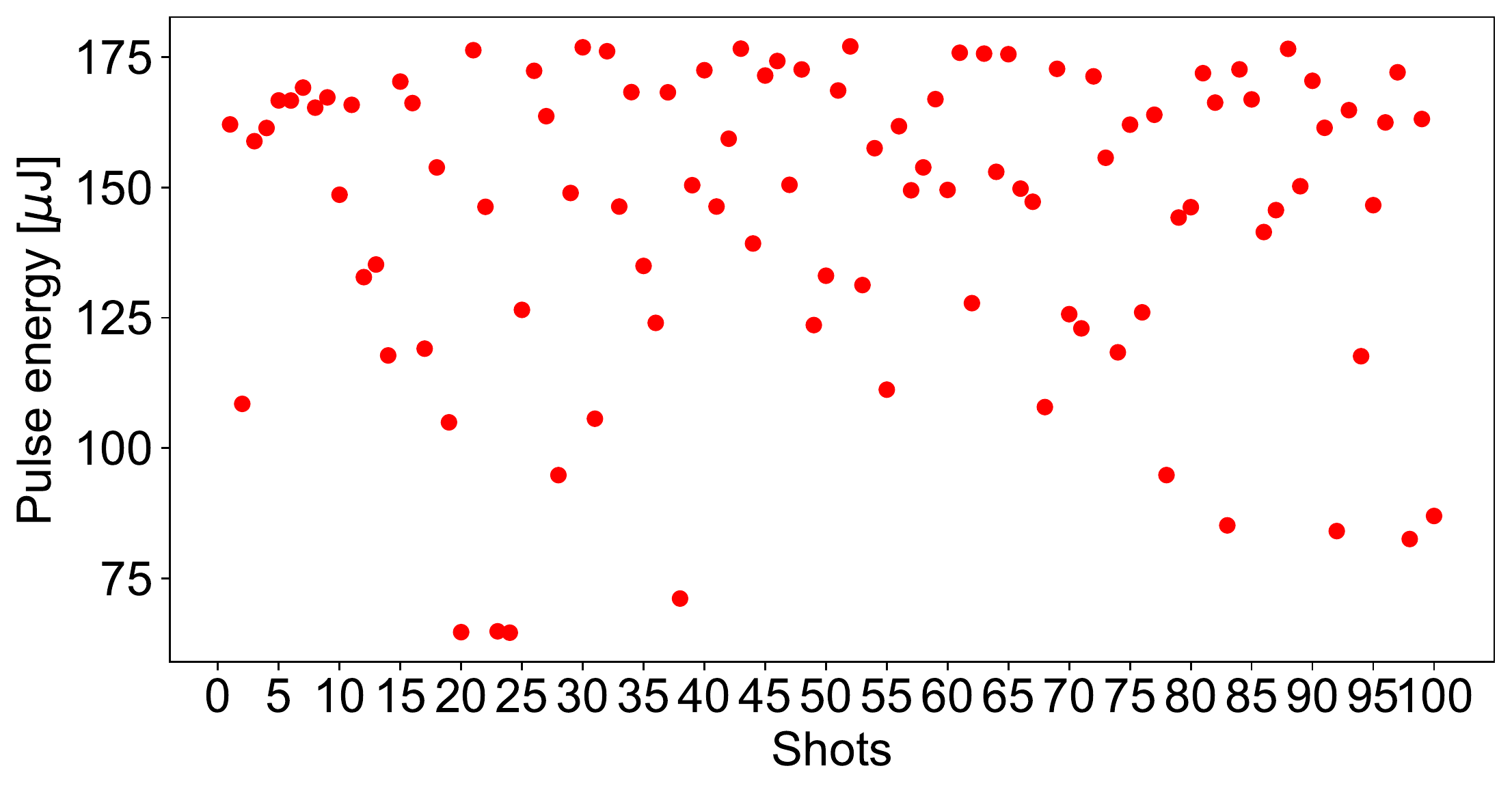}
	\includegraphics[width=0.16\textwidth]{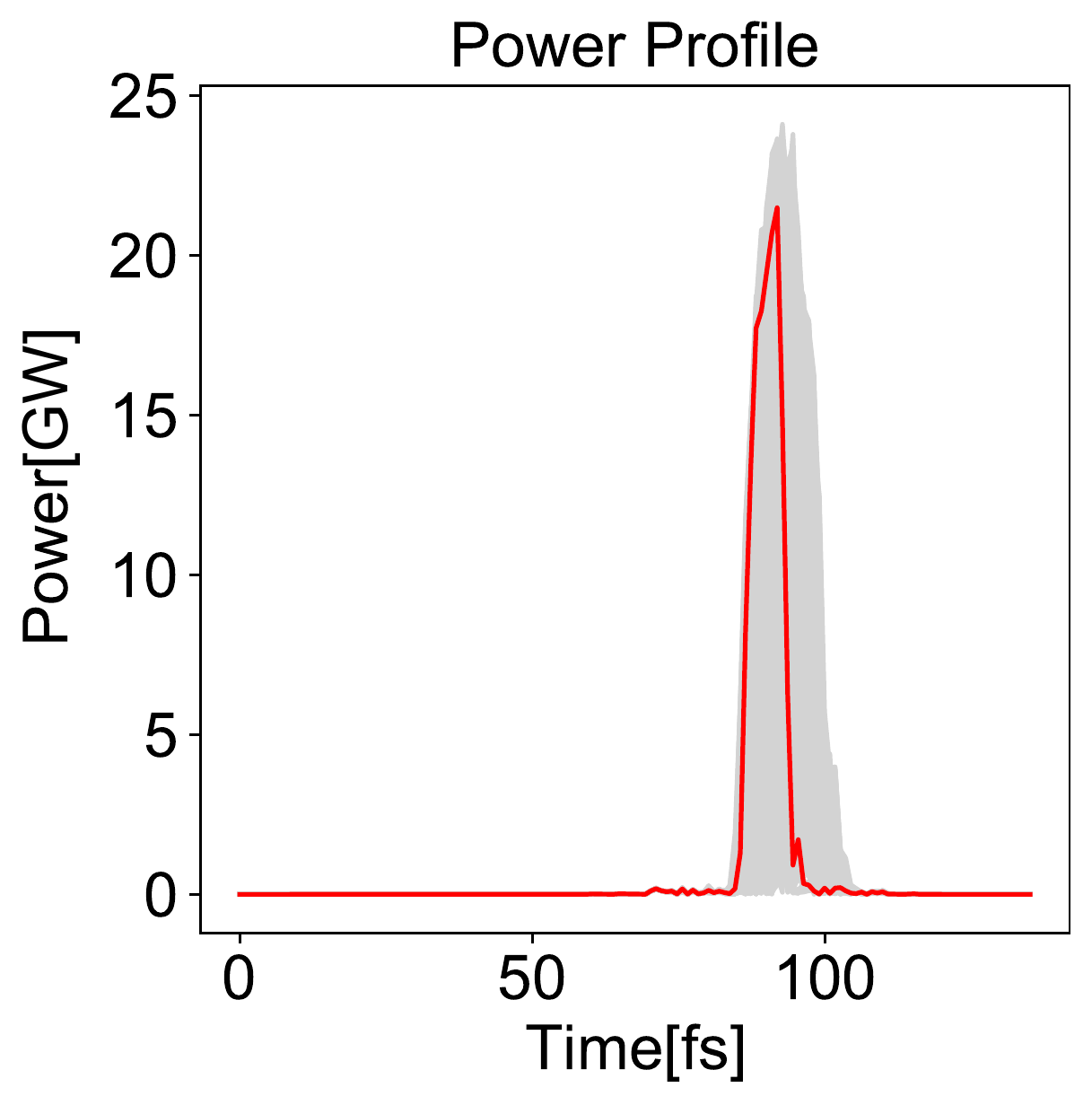}
	\quad
	\includegraphics[width=0.28\textwidth]{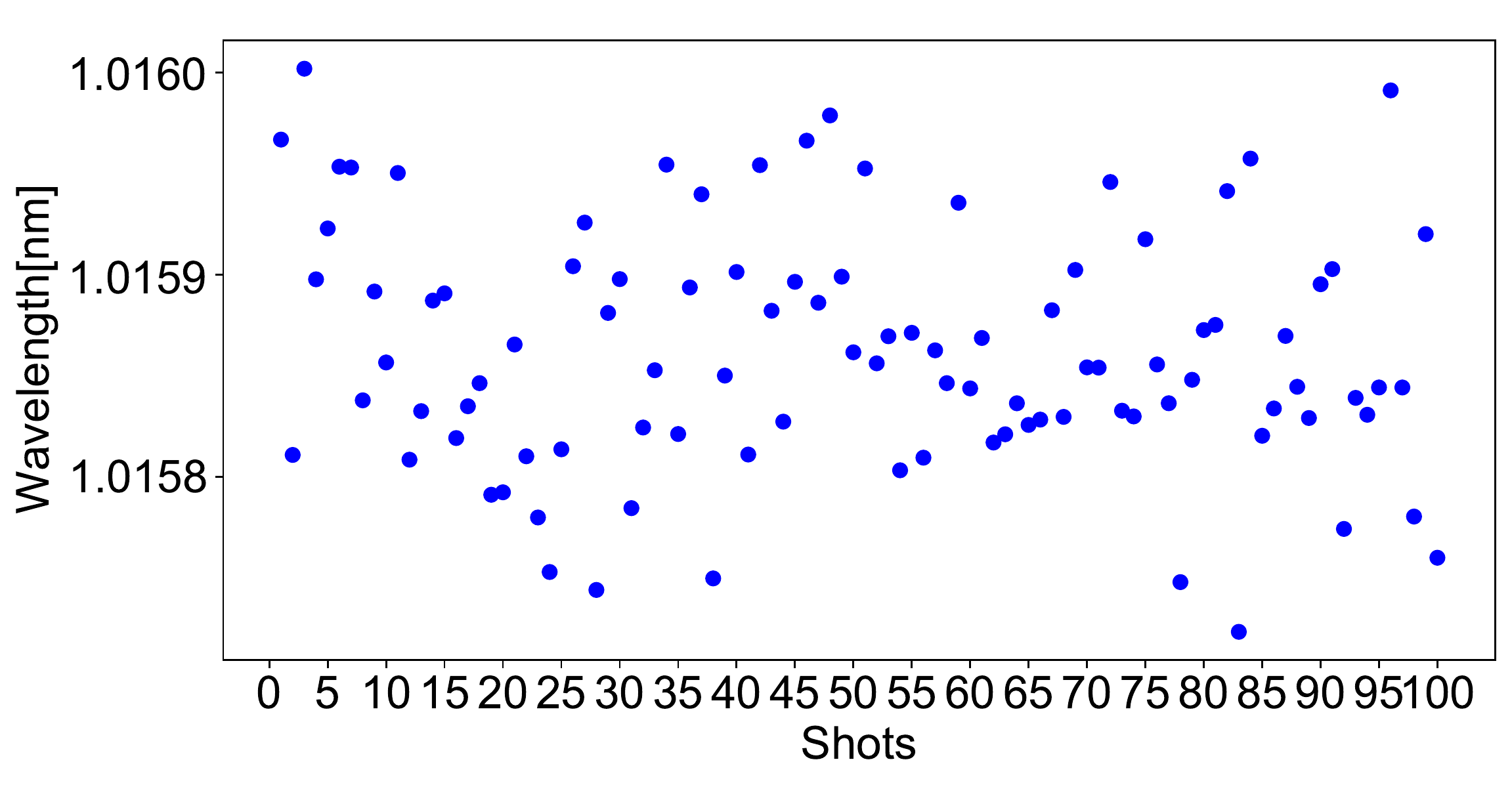}
	\includegraphics[width=0.16\textwidth]{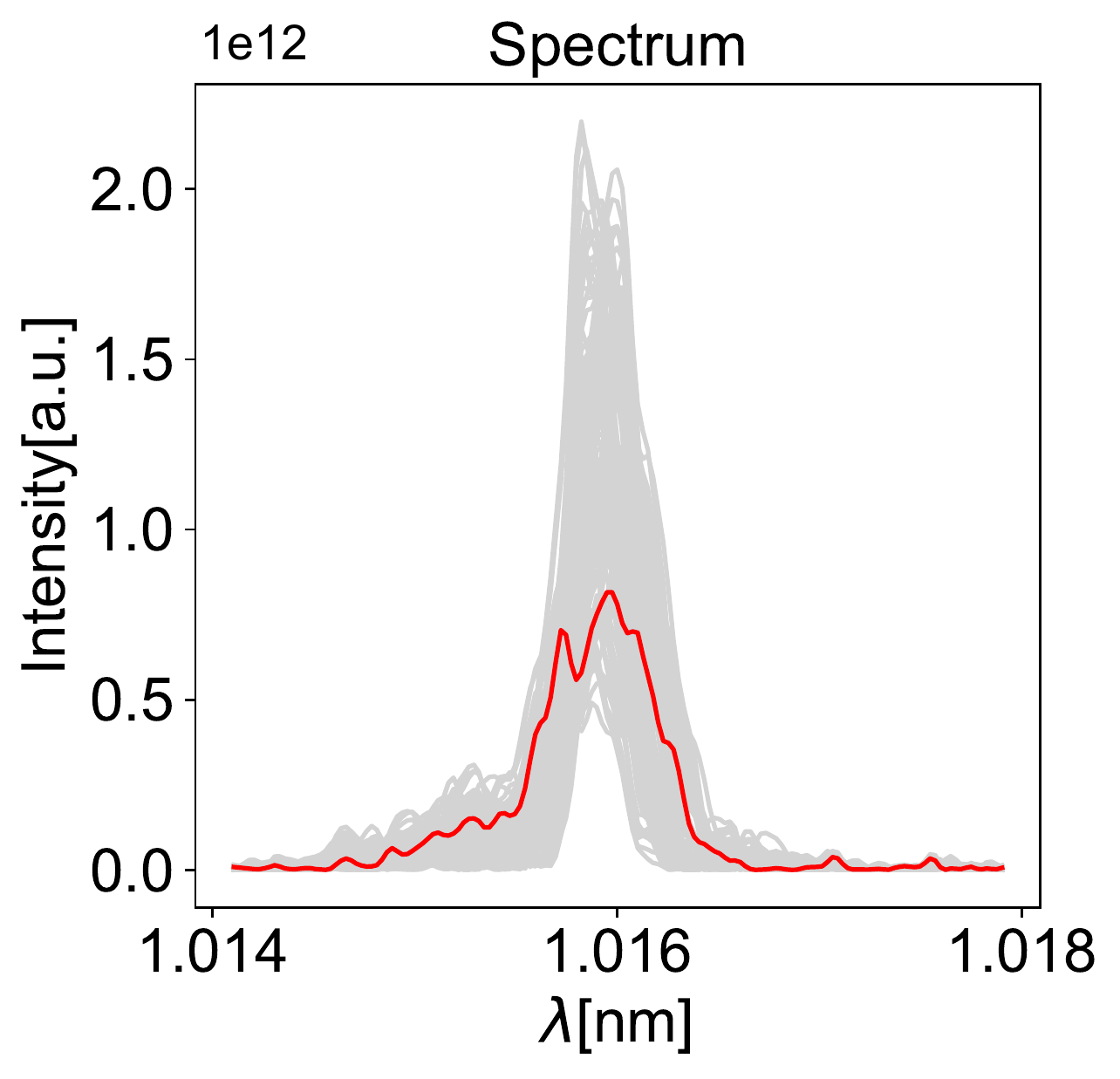}
	\caption{\label{fig:flat1} The 100 shots start-to-end simulations of the second stage HGHG FEL performance at 30 m along the radiators with flat-top current profile.}
\end{figure}

\begin{figure}
    \centering
	\includegraphics[width=0.3\textwidth]{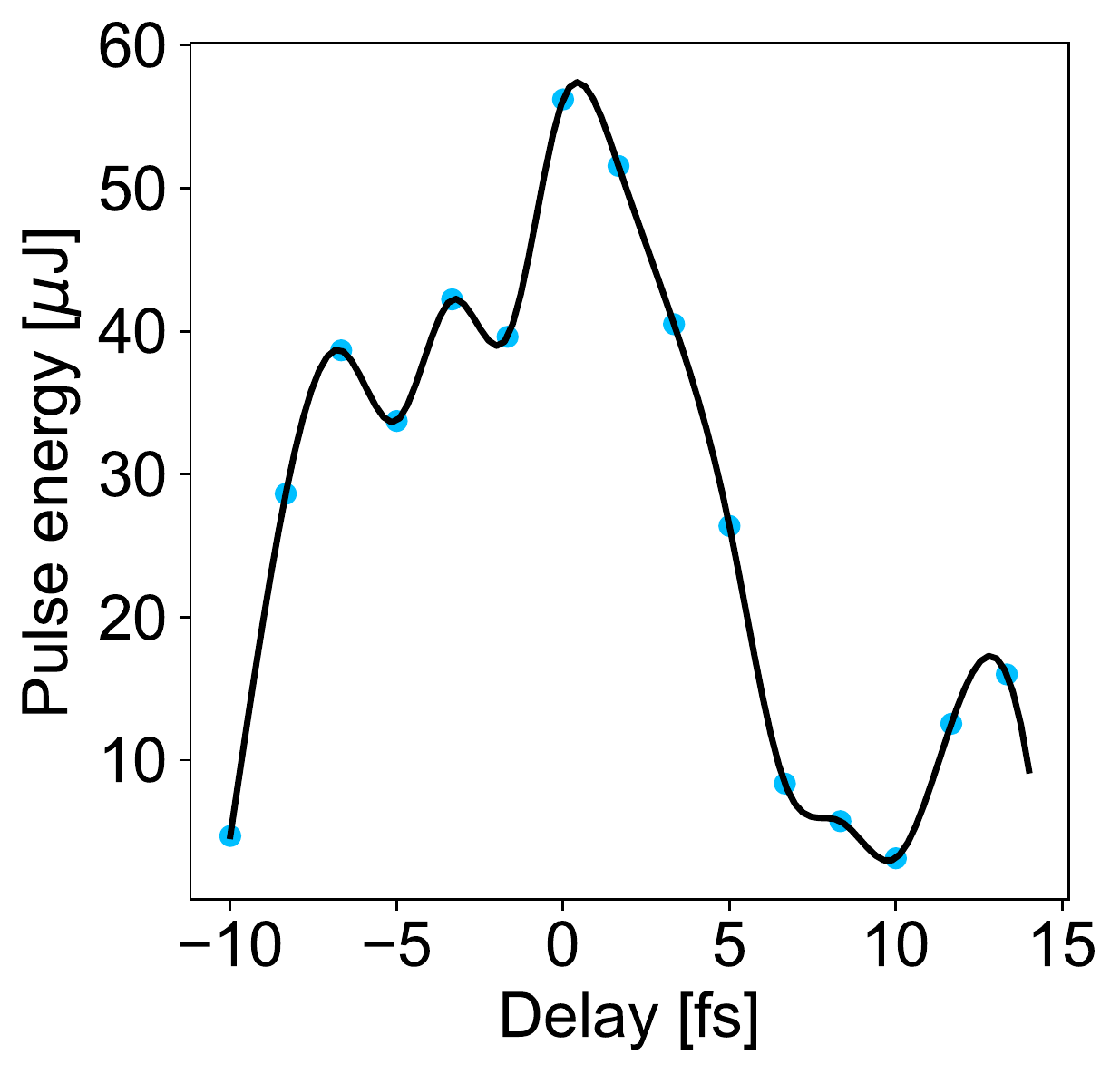}
	\quad
	\includegraphics[width=0.3\textwidth]{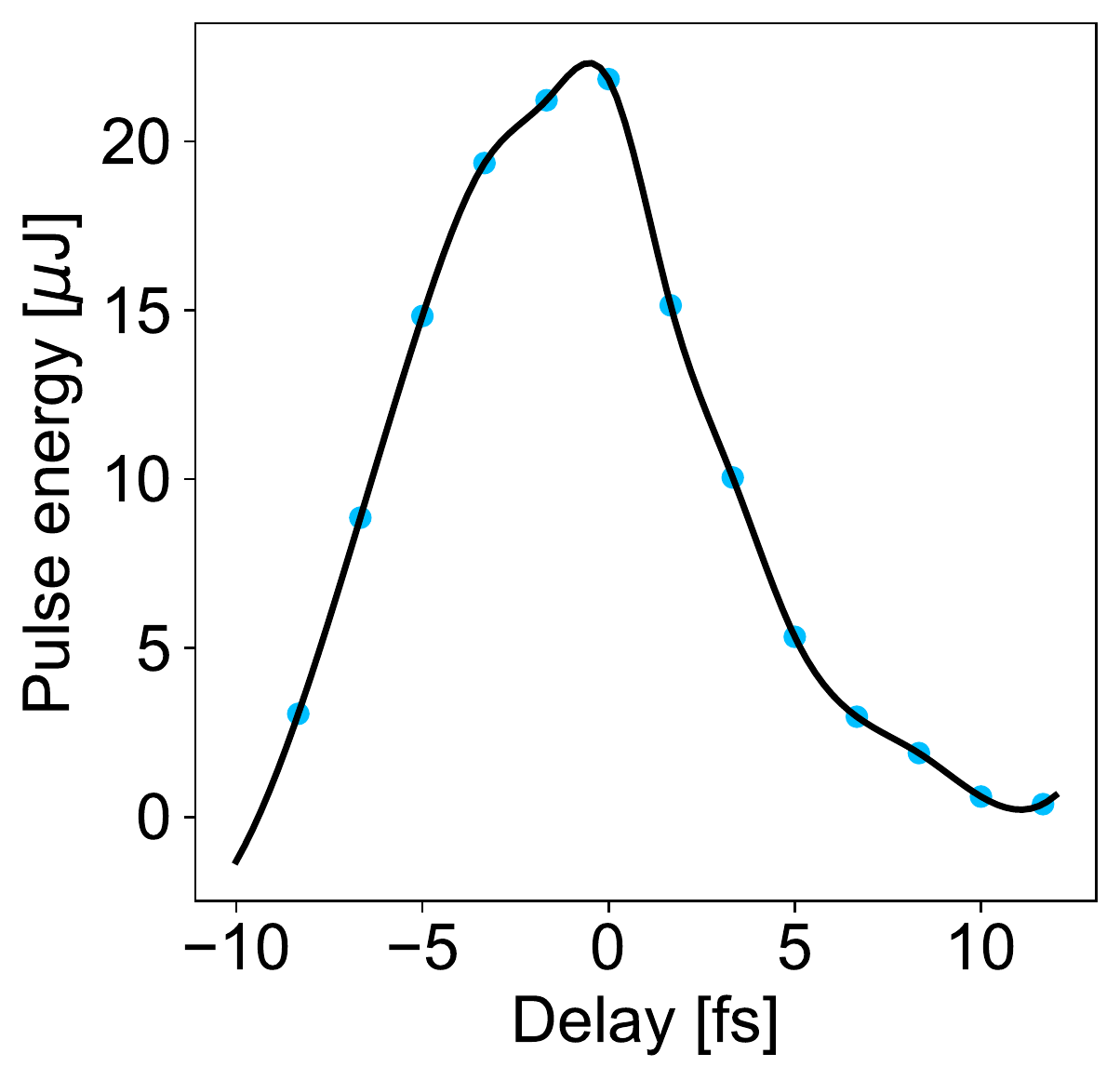}
	\quad
	\caption{\label{fig:jitter1} Sensitivity of the FEL performance against electron beam properties of the second stage HGHG with double-horn (the left) and flat-top (the right) current profiles, respectively. The black line is the interpolation curve. The delay = 0 corresponds to the maximum output FEL pulse energy.}
\end{figure}

\begin{figure}
    \centering
	\includegraphics[width=0.32\textwidth]{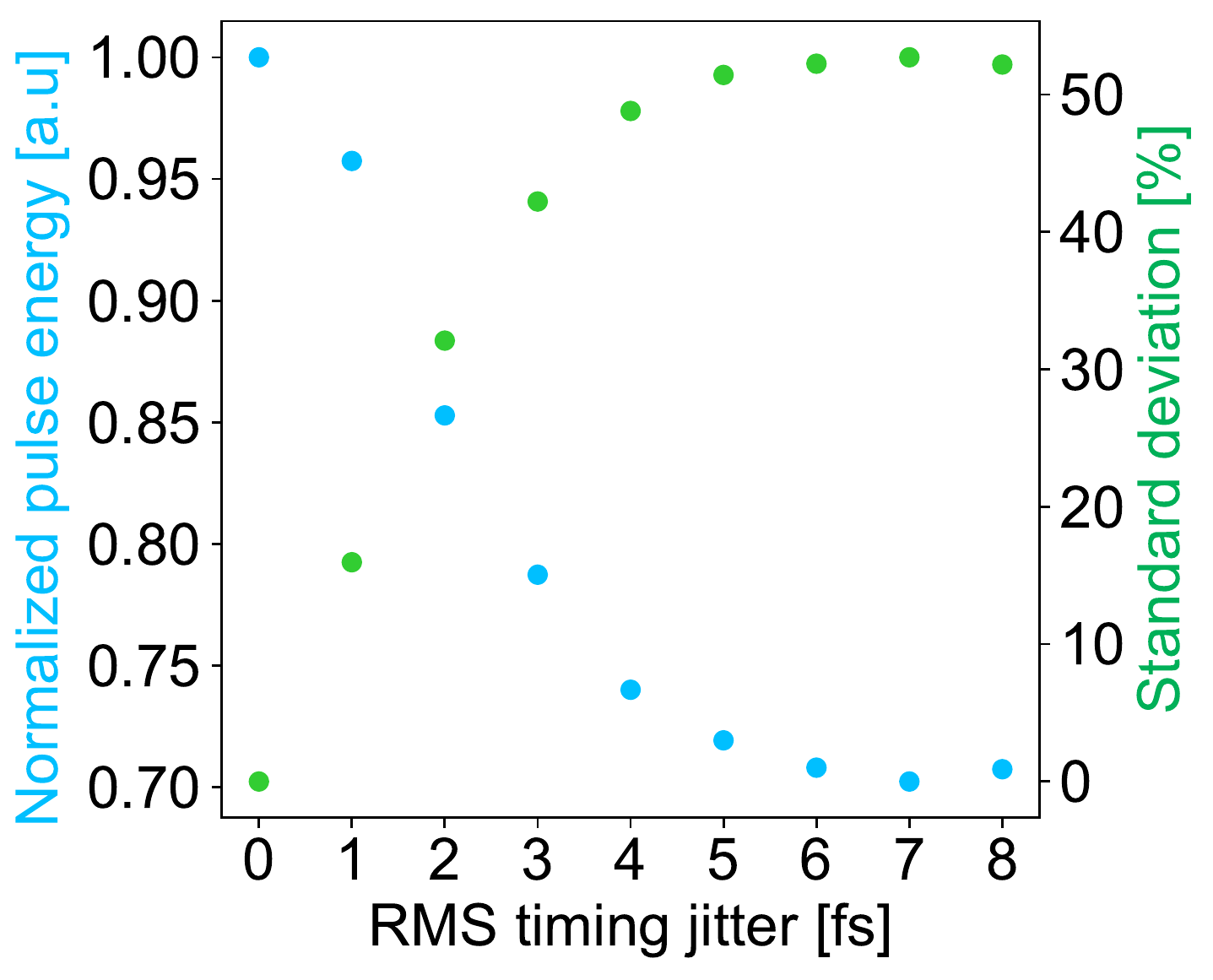}
	\quad
	\includegraphics[width=0.32\textwidth]{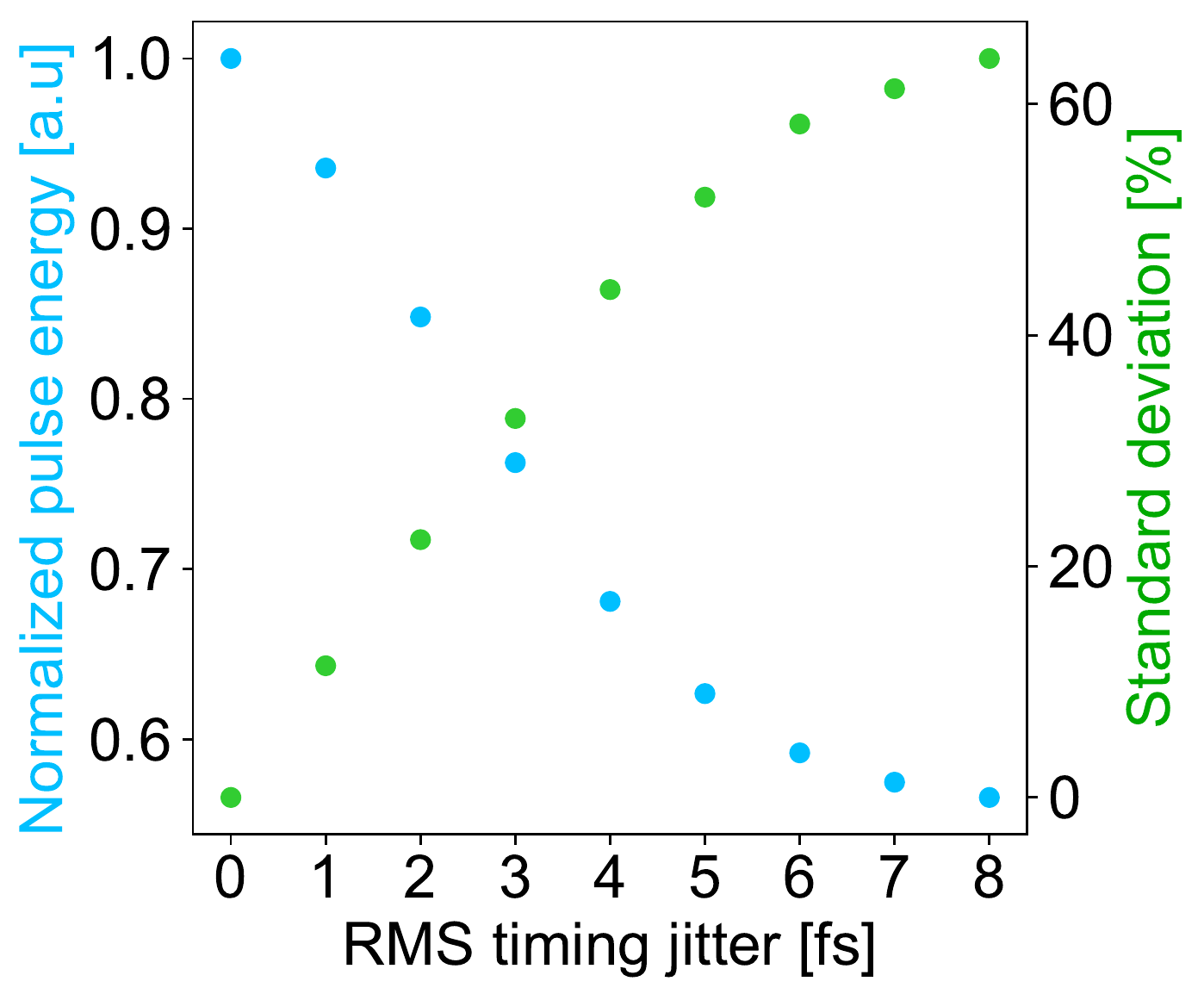}
	\caption{\label{fig:jitter2} Correlations between the pulse energy and RMS timing jitter with double-horn (the left) and flat-top (the right) current profiles, respectively. The blue dots represent the pulse energy normalized by the mean and the green dots indicate the standard deviation over $10^{4}$ samples.}
\end{figure}

\subsection{The second stage HGHG at 20 m and 30 m}
With the “fresh bunch” technique {\cite{Yu1997}}, the bunch tail of 20 fs is manipulated by the external seed lasers and the bunch head continues to interact with the radiation from the first stage in M3 after a time delay of 20 fs corresponding to $R_{56}$ of 14 $\mu$m (see Table~\ref{tab:table1}). The FEL performance of the second stage HGHG at 20 m along the radiators with Gaussian, double-horn and flat-top profiles are shown in Fig.~\ref{fig:HGHG20}, respectively. As the Table~\ref{tab:table2} is illustrated, the mean pulse energy at 20 m with Gaussian, double-horn, and flat-top profiles are 51.08 $\mu$J, 38.90 $\mu$J, and 16.15 $\mu$J, corresponding to standard deviations of 16.71 $\mu$J, 16.40 $\mu$J, and 5.64 $\mu$J, and the fluctuations of 32.72\%, 42.14\%, and 34.95\%, respectively. The FEL pulse energy fluctuations with the flat-top profile of 34.95\% are similar to the Gaussian profile of 32.72\%. In contrast, the pulse length of the double-horn profile is significantly shorter than the others, while the bandwidth is significantly broadened. Because the FEL gain is sensitive to the current profile, which is proportional to $I^{1/3}$ {\cite{Huang2007}}. The slice parameters at the seeding portion of the second stage vary drastically, such as current and energy spread, which can easily destroy the optimized values for seeding, resulting in a shorter pulse length {\cite{Finetti2017}}. Therefore, the flat-top profile with relatively more linearized longitudinal phase space has a more stable output.

In addition, the statistical S2E simulated results of FEL performance at 30 m along the radiators with Gaussian, double-horn, and flat-top profiles are shown in Fig.~\ref{fig:gauss1}, Fig.~\ref{fig:double1}, and Fig.~\ref{fig:flat1}, respectively. Table~\ref{tab:table2} represents that the mean pulse energy at 30 m with Gaussian, double-horn, and flat-top profiles are 225.77 $\mu$J, 229.08 $\mu$J, and 145.37 $\mu$J, corresponding to standard deviations of 24.81 $\mu$J, 22.68 $\mu$J, and 29.46 $\mu$J, and the fluctuations of 11.50\%, 9.90\%, and 20.26\%, respectively. As the interaction between the electron beam and the electromagnetic field deepens, the FEL pulse energy fluctuations with the flat-top profile of 20.26\% become more stable while maintaining the longitudinal coherence, as does the Gaussian profile of 11.50\%. The longitudinal coherence of the double-horn profile is finally destroyed by SASE.

\begin{figure}
    \centering
	\includegraphics[width=0.25\textwidth]{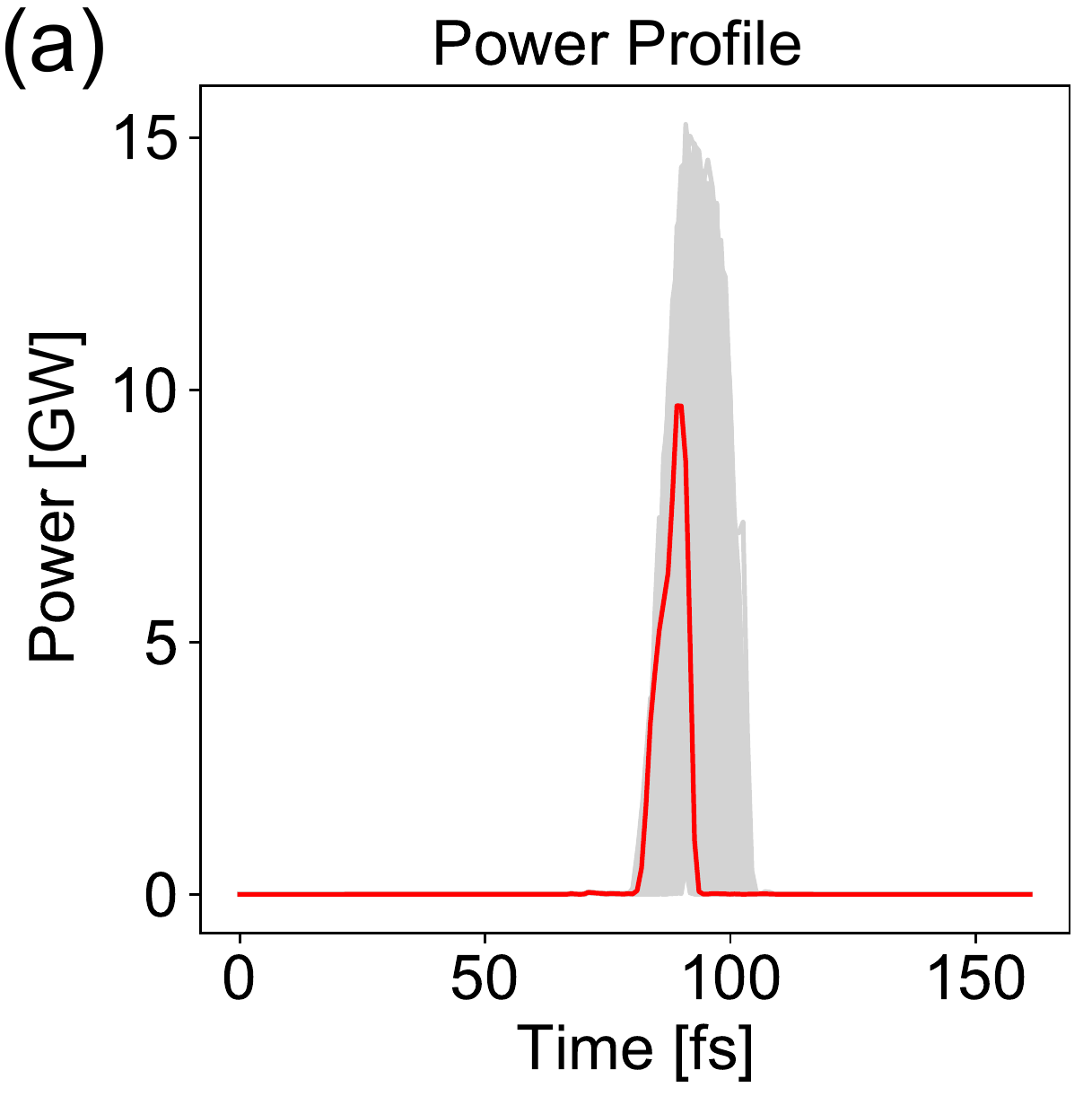}
	\quad
	\includegraphics[width=0.25\textwidth]{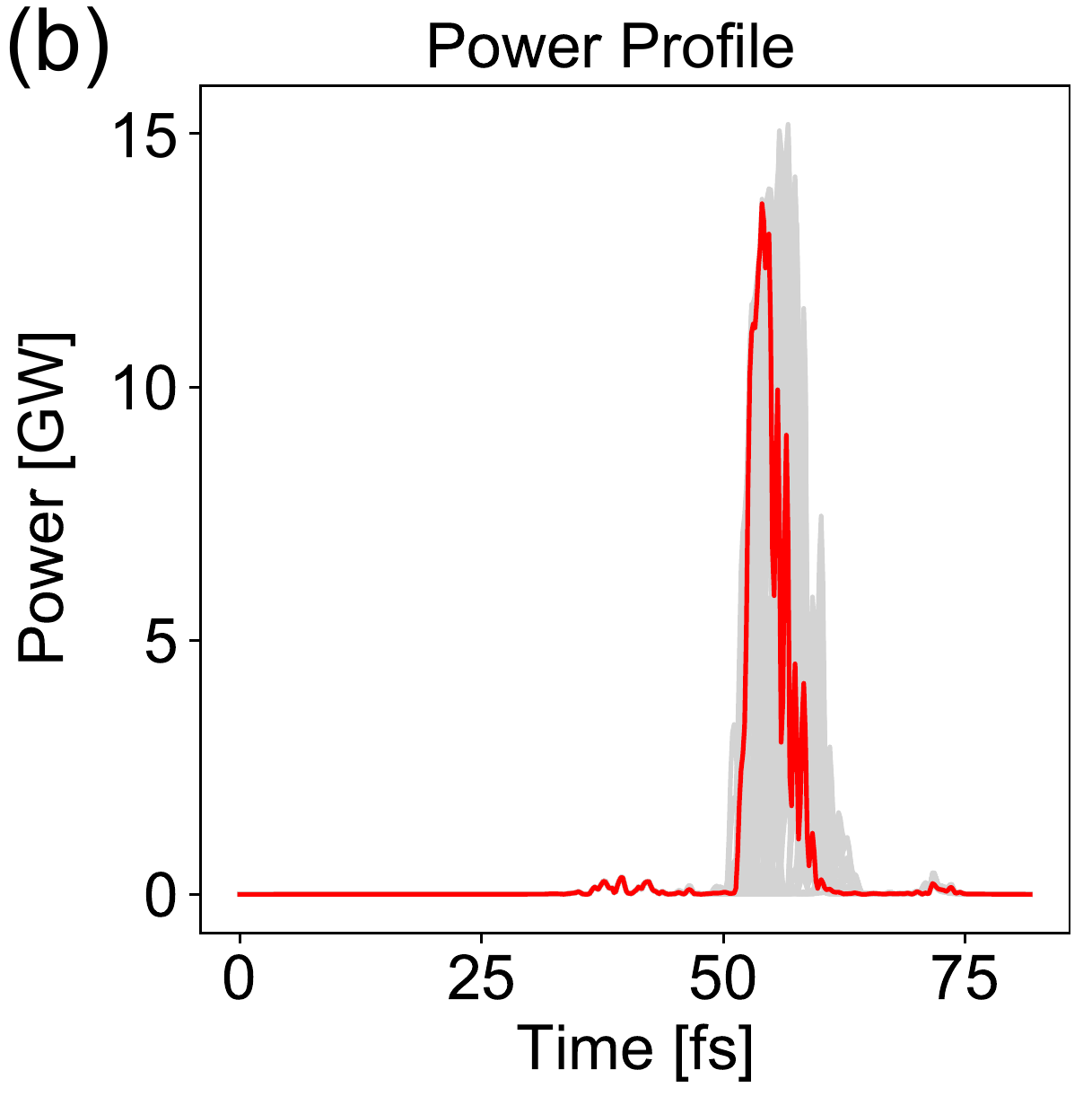}
	\quad
	\includegraphics[width=0.25\textwidth]{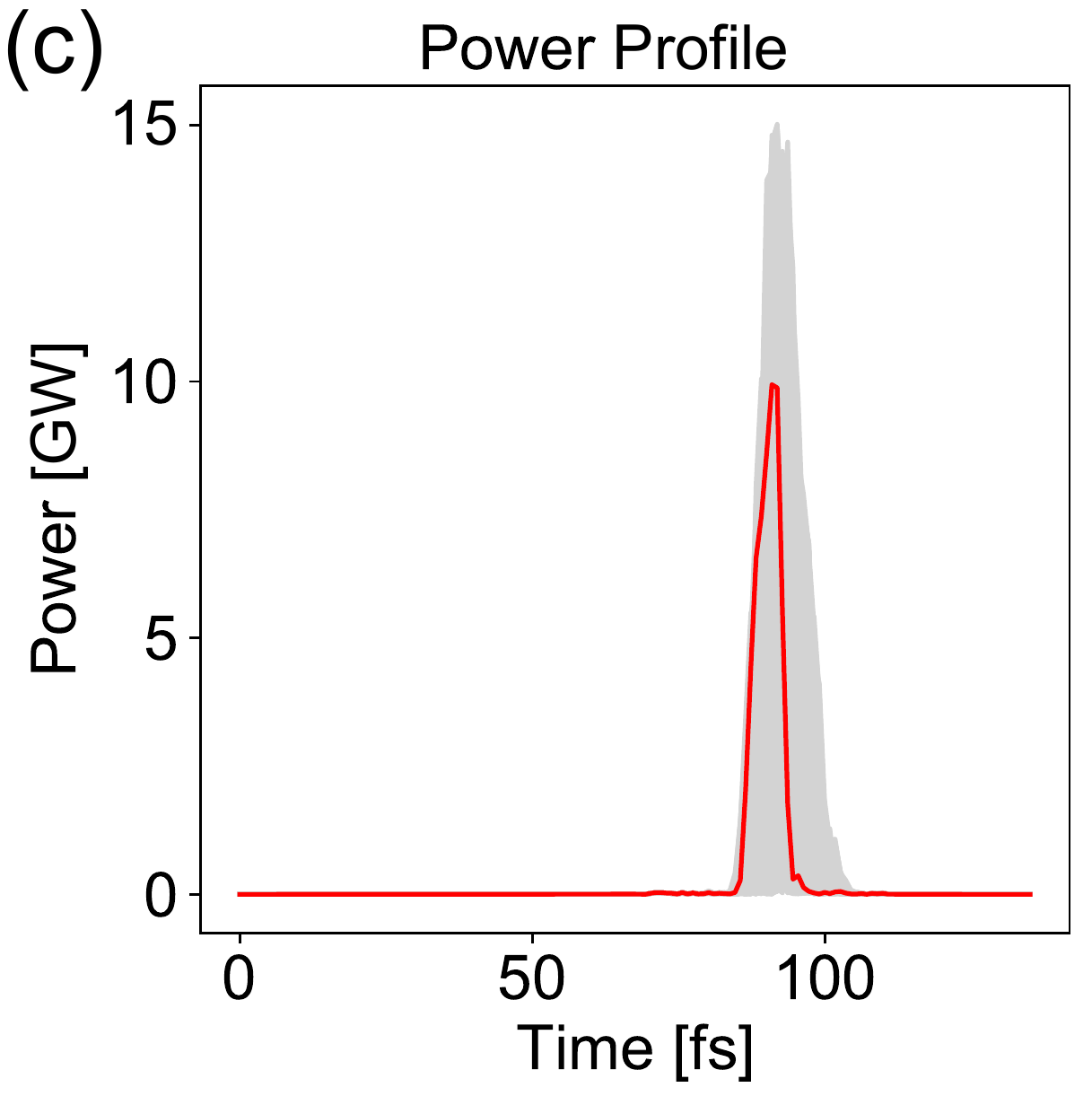}
	\caption{\label{fig:15GW_power} 100 shots of final FEL power profiles with the same maximum peak power of 15 GW and different current profiles. (a) Gaussian profile; (b) Double-horn profile; (c) Flat-top profile.}
\end{figure}

\begin{figure}
    \centering
	\includegraphics[width=0.3\textwidth]{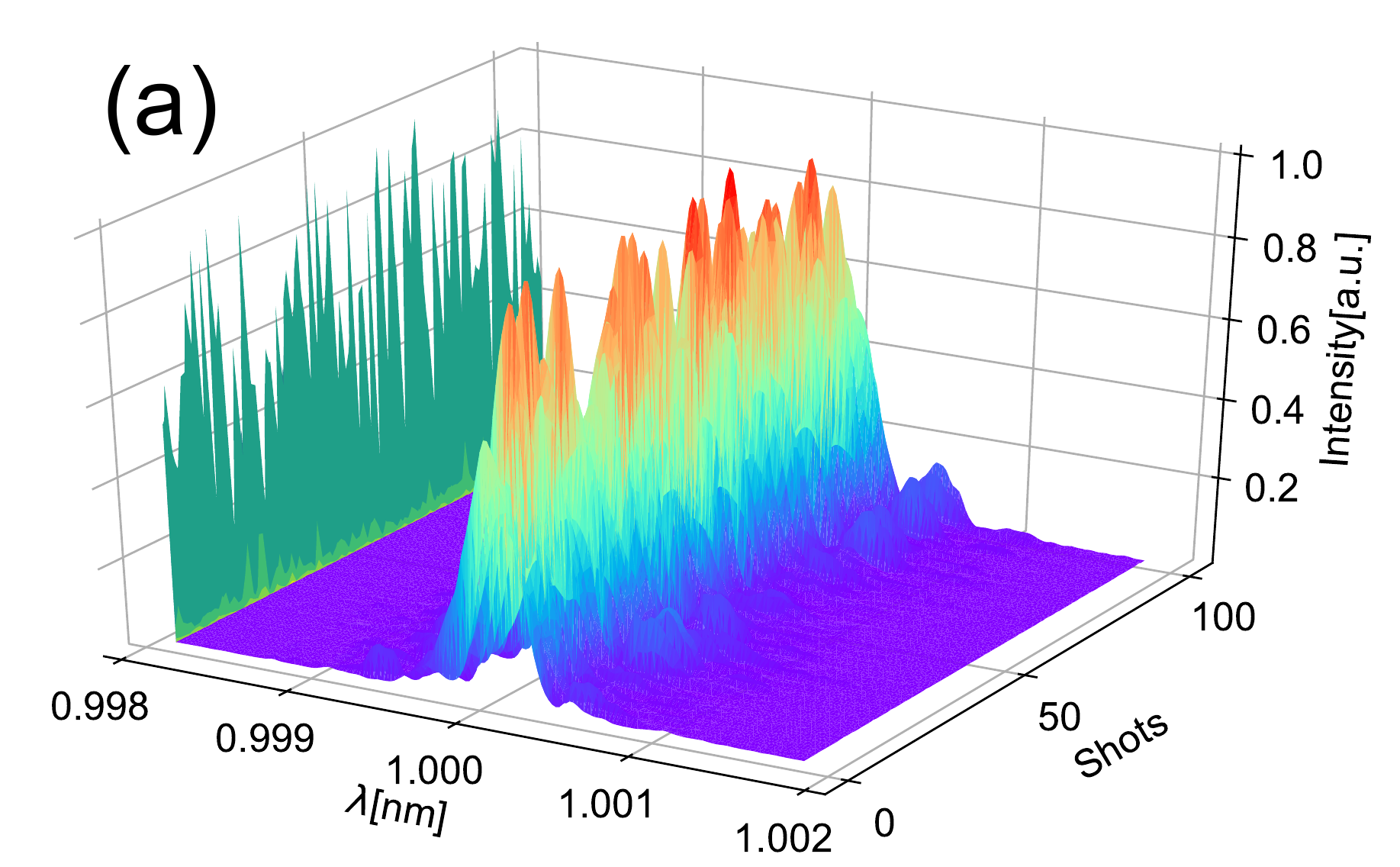}
	\quad
	\includegraphics[width=0.3\textwidth]{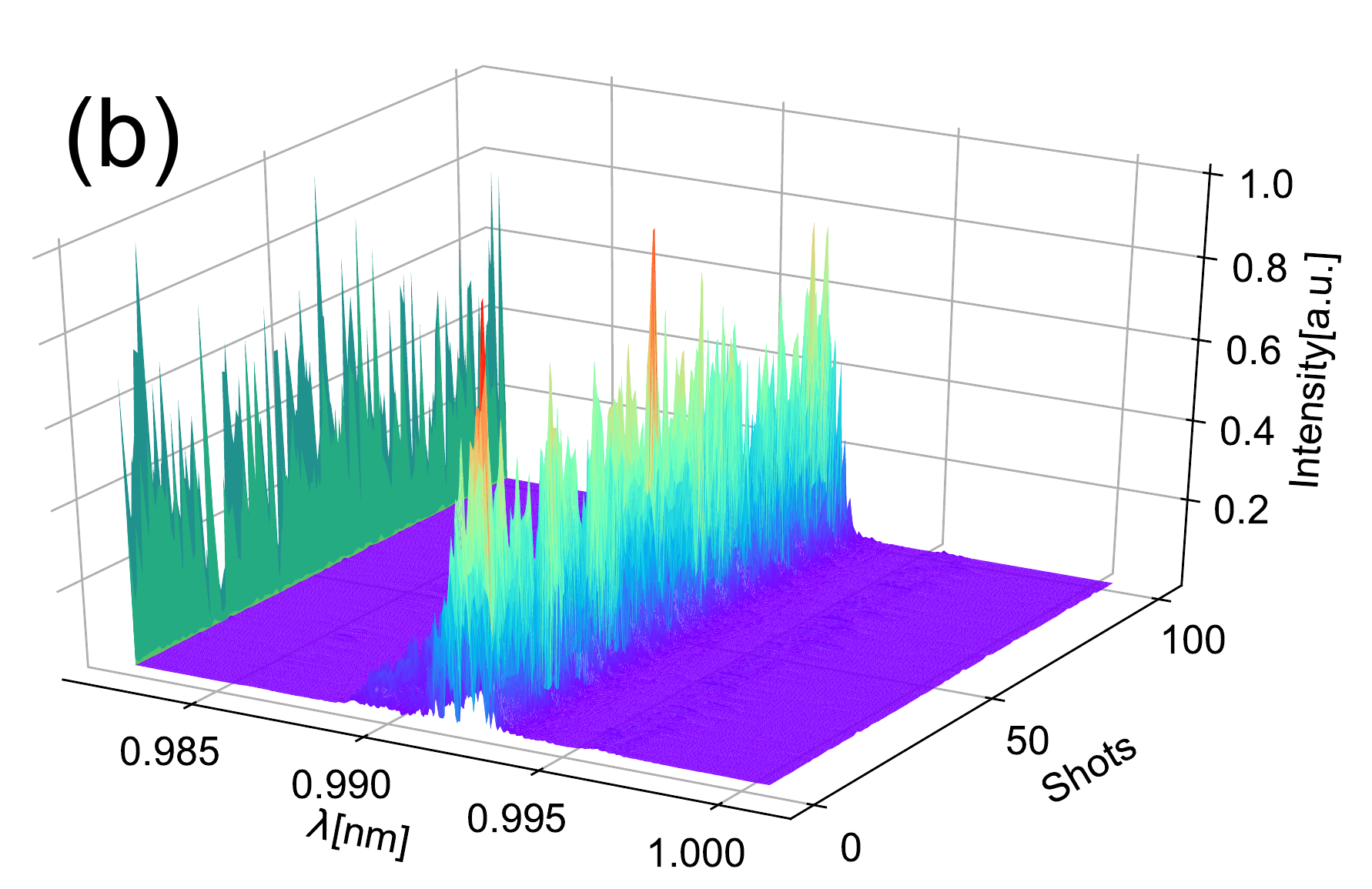}
	\quad
	\includegraphics[width=0.3\textwidth]{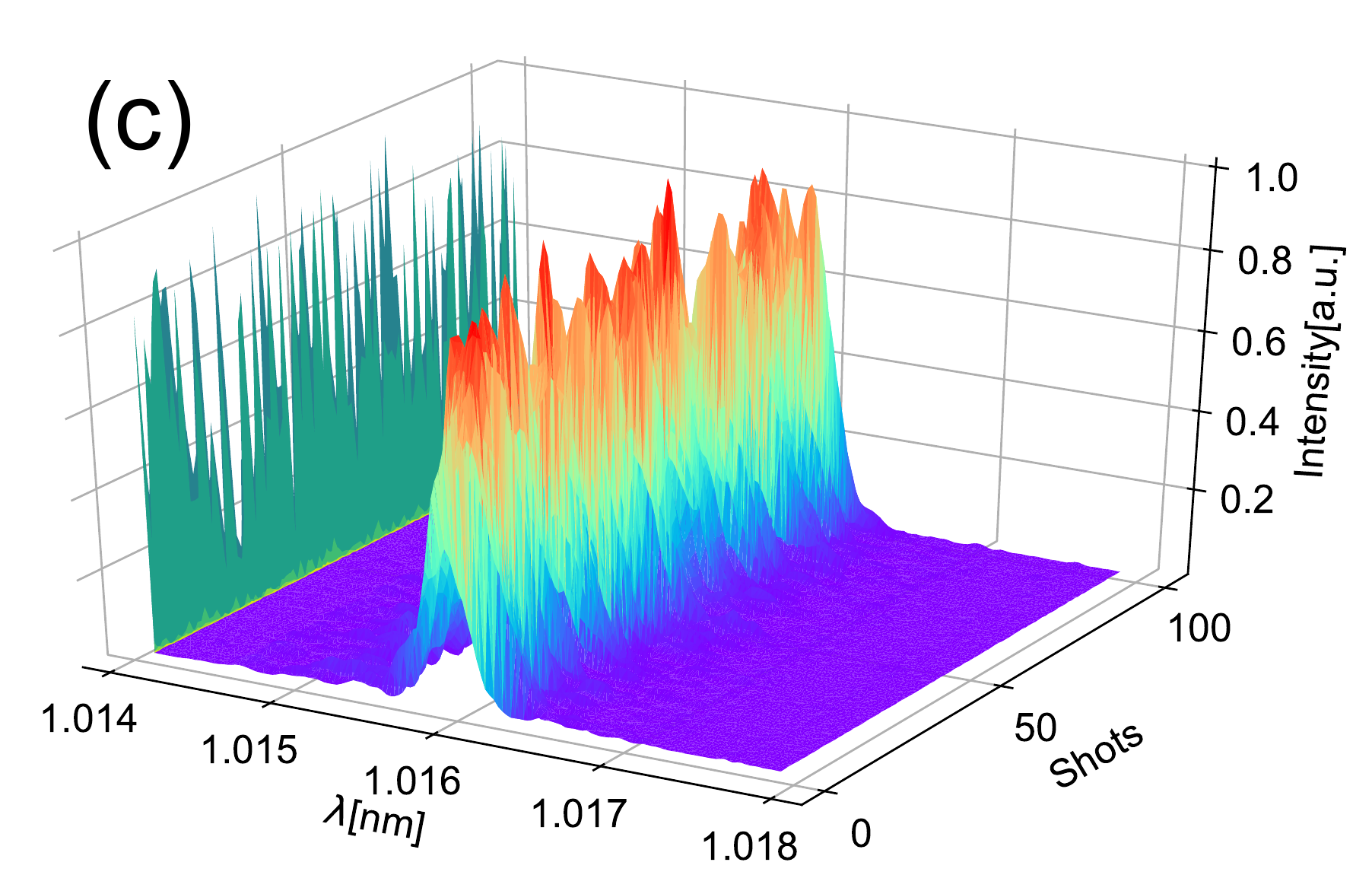}
	\caption{\label{fig:spectra} 100 shots of final FEL spectra with the same maximum peak power of 15 GW and different current profiles. (a) Gaussian profile; (b) Double-horn profile; (c) Flat-top profile.}
\end{figure}

\begin{table*}
    \centering
	\caption{\label{tab:table3}The statistical simulated results of the FEL performance of the first and second stages with gaussian profile, double-horn profile, and flat-top profile.}
	\begin{tabular}{cccc}
        \toprule
		Current profile&Gaussian@23.5m&Double-horn@21m&Flat-top@26m\\ \midrule
		Mean pulse energy ($\mu$J)&108.01&41.07&61.91\\
		Standard deviation ($\mu$J)&27.68&17.06&18.05\\
		Fluctuation (\%)&25.63&41.56&29.16\\
		Mean bandwidth (\%)&0.027&0.132&0.030\\
		Standard deviation (\%)&0.0051&0.0212&0.0047\\
		Fluctuation (\%)&18.69&16.08&15.50\\
		\bottomrule
	\end{tabular}
\end{table*}

\section{\label{sec:level4}Sensitivity analysis of the timing jitter}
To further investigate the feasibility of the cascaded EEHG-HGHG scheme with different current profiles, the sensitivity of the FEL performance of the second stage HGHG against electron beam properties is shown in Fig.~\ref{fig:jitter1}. The double-horn and flat-top correspond to a scan of about 25 fs, thus the black curves obtained by interpolation show that the pulse energy is a function of the time delay. The delay = 0 represents that the maximum output FEL pulse energy can be obtained and considered to be the absence of relative timing jitter.

Meanwhile, the interpolation curves can evaluate FEL performance with a given RMS timing jitter as a prediction model {\cite{yang:ipac}}. With this model, one can investigate timing jitter effects on FEL stability under different current profiles by sampling. As presented in Fig.~\ref{fig:jitter2}, the correlations between the pulse energy and RMS timing jitter with Gaussian, double-horn, and flat-top profiles, respectively, are obtained, where each RMS values correspond to $10^{4}$ samples. The larger RMS timing jitter means the larger reduction of the mean pulse energy and deviations. As we discussed above, RMS timing jitter of 3 fs is added to S2E numerical simulations, thus the predicted FEL pulse energy fluctuations of double-horn and flat-top profiles are 42.62\%, and 33.45\%, respectively, which are remarkably close to the results of 42.14\%, and 34.95\% (see Table~\ref{tab:table2}) obtained from S2E numerical simulations. Therefore, this simple method can be applied to the impact of the relative timing jitter with different RMS values under different current profiles and reduce the number of S2E numerical simulations.

More importantly, to further compare the effect of relative timing jitter on the final output FEL performance and remove the differences in the FEL gain process brought by the differences in optimized parameters under different cases, the FEL spectra of the corresponding radiator positions with the same maximum peak power are selected for analysis. The final FEL power profiles with Gaussian, double-horn, and flat-top profiles at 23.5 m, 21m, and 26 m along the radiators are shown in Fig.~\ref{fig:15GW_power}, respectively. The final FEL spectra are illustrated in Fig.~\ref{fig:spectra} and the statistical simulated results are listed in Table~\ref{tab:table3}. The mean pulse energy of the flat-top profile is 61.91 $\mu$J larger than the 41.07 $\mu$J of the double-horn profile but both are lower than the 108.1 $\mu$J of the Gaussian profile, while the pulse energy stability is close to that of the ideal Gaussian, 29.16\% and 25.63\%, respectively, better than the 41.56\% of the double-horn profile; the mean bandwidth of the double-horn profile is $1.3 \times 10^{-3}$ due to the energy chirp in the second-stage HGHG {\cite{WANG201456}}, which is nearly 4.5-fold larger than that of the Gaussian and flat-top profile of $2.7 \times 10^{-4}$ and $3.0 \times 10^{-4}$, but the mean bandwidth (FWHM) fluctuations is similar for all three corresponding to 16.08\%, 18.69\%, and 15.50\%. The sensitivity analysis of timing jitter indicates that the flat-top current profile performs significantly in improving longitudinal coherence for the two-stage seeded FEL.

\section{\label{sec:level5}Conclusions}
In this paper, the feasibility of the cascaded EEHG-HGHG scheme at SHINE is designed and optimized for generating coherent radiation with a harmonic up-conversion number of 270. The comparison of the FEL performance at different stages and different radiator positions is discussed by intensive start-to-end numerical simulations under double-horn and flat-top profiles, respectively. With an RMS timing jitter of 3 fs, the flat-top profile is similar to the ideal Gaussian profile in terms of stability, which is significantly better than the double-horn; in particular, the first stage EEHG with the flat-top profile is the more stable than the others. Meanwhile, the interpolation curves of RMS timing jitter and pulse energy to estimate the effects of timing jitter under different RMS values can significantly reduce numerical simulations. In addition, the sensitivity analysis by comparing the final FEL output with the same maximum peak power of 15 GW under different current profiles shows that the flat-top profile has no significant advantage on the pulse energy stability. However, although the bandwidth stability with different current profiles is similar, the longitudinal coherence of the flat-top profile is significantly better than the others.

Overall, the FEL stability with the flat-top profile is a favorable candidate for the cascaded EEHG-HGHG scheme, which can relax the timing jitter requirement and generate coherent FEL pulses. There are still numerous practical issues in implementing two-stage schemes in tens of femtoseconds electron beam, such as RMS timing jitter lower than 3 fs, transverse instability of 10-GW level seed lasers, and IBS {\cite{Piwinski1988}} and CSR {\cite{Huang2002}} effects in EEHG scheme as harmonic increases. However, seeded FELs based on self-modulation scheme {\cite{Yan2021}} or optical cavity {\cite{Petrillo2019,Ackermann2020,Paraskaki2021}} have the potential to reduce the requirements for state-of-art laser system and thus obtain high-repetition-rate fully coherent soft X-ray FEL pulses. In the future, the physical design of the externally seeded FEL at SHINE needs to be further optimized.

\section{Acknowledgement}
The authors would like to thank Z. Gao and N. Huang for their helpful discussions. This work was supported by the National Key Research and Development Program of China (2018YFE0103100), the National Natural Science Foundation of China (12125508, 11935020), Program of Shanghai Academic/Technology Research Leader (21XD1404100), and Shanghai Pilot Program for Basic Research - Chinese Academy of Science, Shanghai Branch (JCYJ-SHFY-2021-010).

\bibliography{manuscript}

\end{document}